%% file: orbit_resub.tex
\documentclass[aps,pre,reprint,notitlepage,onecolumn,groupedaddress]{revtex4-1}
\usepackage{epsfig,graphics,amssymb,amsmath,amsfonts,color,subeqnarray}
\usepackage[sfdefault=cmbr]{isomath} 
\usepackage{subcaption}
\usepackage[colorlinks=true,linkcolor=blue]{hyperref}

\input{definitions}

\begin{document}
\title{Jeffery orbits in shear-thinning fluids}
\author{S. Arman Abtahi}
\affiliation{Department of Mechanical Engineering and Institute of Applied Mathematics, 
University of British Columbia,
Vancouver, BC, V6T 1Z4, Canada}
\author{Gwynn J. Elfring}\email{gelfring@mech.ubc.ca}
\affiliation{Department of Mechanical Engineering and Institute of Applied Mathematics, 
University of British Columbia,
Vancouver, BC, V6T 1Z4, Canada}
\date{\today}
\begin{abstract}
{
We investigate the dynamics of a prolate spheroid in a shear flow of a shear-thinning Carreau fluid. The motion of a prolate particle is developed analytically for asymptotically weak shear thinning and then integrated numerically. We find that shear-thinning rheology does not lift the degeneracy of Jeffery orbits observed in Newtonian fluids but the instantaneous rate of rotation and trajectories of the orbits are modified. Qualitatively,  shear thinning  has a similar effect to elongating the particle in a Newtonian fluid. The period of rotation increases as the particle slows down more when aligned with the flow due to a reduction of shear stresses. Unlike Jeffery orbits in Newtonian fluids, in shear-thinning fluids the period of orbits depends on the specific trajectory (or initial orientation of the particle).
}
\end{abstract}
\maketitle

\section{Introduction}
Pastes \cite{Coussot_2005}, paints \cite{Eley2005}, pulps \cite{clark1985pulp}, and ceramics \cite{Tsetsekou_2001}, among many other chemical substances \cite{Aust_2011}, contain suspensions of (anisotropic) particles. Suspensions of particles in complex fluids are widely used in the petroleum industry, from drilling muds used to drill wells \cite{hamed09}, to viscoelastic carrier fluids that carry proppant particles used in hydraulic fracturing \cite{barbati16}.  Biological fluids also carry nutrition \cite{Hardacre2018}, Bacillus species \cite{Simon2011}, and other mobile micro-organisms of mostly rod shape \cite{Saverio2015, Dasgupta_2017}.  These are but a few examples that demonstrate the importance of understanding the underlying dynamics and rheology of particles suspended in fluid flows. As we will highlight below, this area has a long literature when the suspending fluid is Newtonian, but our understanding is much more basic when the fluid is non-Newtonian. 

The presence of particles in a fluid can dramatically change its rheology. \citet{Einstein1906,Einstein1911} calculated an effective viscosity for a dilute suspension of non-colloidal hard spheres and showed that the effective viscosity of the suspension increases linearly with the volume fraction of spheres. For higher volume fractions, hydrodynamic (and contact) interactions become more significant. Suspensions generally shear thin with increasing the shear \cite{van_der_Werff_1989} while beyond a threshold exhibit shear thickening or even jamming \cite{Wagner_2009}. If the particles are anisotropic, then even small deviations from spherical geometry can significantly impact the rheology even for dilute suspensions \citep{Leal_1972}. One reason for this is because elongated or rod-like particles tend to align with the flow thus changing the properties of the fluid under shear. Therefore, understanding the dynamics of even a single particle in shear is important to determine the rheology of a suspension of anisotropic particles.

The dynamics of elongated particles, for example axisymmetric ellipsoids and spheroids, display significantly more complicated behaviour than spherical particles under shear \cite{Oberbeck1876,Edwardes1893,Jeffery1922,Chwang1974}. At zero Reynolds number, spheroids and long slender bodies in shear flow undergo a periodic motion. A century ago, \citet{Jeffery1922} solved the motion of a neutrally buoyant ellipsoid of revolution in a simple uniform shear flow in the absence of inertial and Brownian forces. He found that the particle's axis of revolution rotates on infinitely many degenerate periodic orbits called ``Jeffery orbits". Jeffery's solution is degenerate in the sense that the orientation of the body at long times depends on its initial orientation. Jeffery suggested that this degeneracy would be lifted by inertia and speculated that the particle would evolve to an orbit corresponding to the minimum mean energy dissipation. A year after Jeffery's calculations, Taylor \cite{Taylor1923} experimentally showed that an ellipsoid of revolution in a simple shear flow drifts through the continuous family of Jeffery orbits until the ellipsoid is rotating in a final preferred orbit. A prolate spheroid, after approximately 180 complete revolutions, would settle into a \textit{log-rolling} final orbit, rotating perpendicular to the shear plane so that its long axis is parallel to the vortex direction. In contrast, an oblate spheroid, after approximately 40 revolutions, would assume a \textit{tumbling} orbit in which its axis of revolution is in the shear plane and rotates with variable angular velocity. In other words, Taylor confirmed Jeffery's minimum energy hypothesis for spheroidal particles within the range of his experiments. \citet{Harper1968} later showed theoretically and experimentally that the preferred constant orbit for a dumb-bell shaped body corresponds to maximum dissipation (tumbling in the flow-shear plane). \citet{DING_2000} showed through direct numerical simulations  that the period of a prolate spheroid diverges for higher Reynolds numbers as the particle remains motionless when it is nearly aligned with the flow direction. Recently, Einarsson \textit{et al.} \cite{Einarsson2015a, Candelier2015, Einarsson2015b, Ros2015, Einarsson2016, Byron2015} have shown theoretically that in the limit of weak flow and particle inertia, the degeneracy of Jeffery orbits is indeed lifted. The first effects due to weak inertia cause a prolate spheroid in simple shear to drift to a stable tumbling limit cycle, whatever the initial condition \cite{Einarsson2015a}.

Bretherton \cite{Bretherton1962} investigated the motion of a particle of more general shape in shear flow and in the presence of rigid boundaries. He showed that axisymmetric particles follow  Jeffery's  equation of motion if the aspect ratio is replaced with an approximate effective aspect ratio. The theoretical work of \citet{Hinch1979} on non-axisymmetric ellipsoids in shear flow shows that even a small deviation from axisymmetric geometry results in profound changes in the nature of the orbit. Consequently, theoretical models quantifying real solutions of particles, on the assumption of axisymmetric particles, can be inaccurate. The motion of a non-axisymmetric ellipsoid has two periodic parts called ``doubly periodic'' tumbling: a rapid rotation similar to Jeffery orbits and a slower drift in the orbits. Recently, \citet{masoud2013} showed numerically that porous ellipsoids follow Jeffery orbits (of impermeable ellipsoids) to very good approximation.

As noted above, most prior studies have focused on Newtonian fluids, but for suspensions in non-Newtonian fluids the fundamental building blocks governing the rheology are still being developed. Non-Newtonian carrier fluids can produce qualitative changes in the rheological behaviour of suspensions in comparison with Newtonian fluids, as an example, recent studies have calculated modifications of the Einstein viscosity for dilute suspensions of spheres in weakly nonlinear viscoelastic fluids \citep{Einarsson_2018}, and shear-thinning fluids \citep{Datt2018}. 

The dynamics of individual particles in flows can also be substantially modified by complex fluid rheology. Early work focused on the dynamics of spherical particles in weakly nonlinear viscoelastic fluids, and much of that work is summarized in the  wonderful  review by \citet{Leal1980}. The dynamics of anisotropic particles in shear flows of viscoelastic fluids has also received attention. In experimental work, \citet{Saffman_1956} reported that spheroidal particles deviate from Jeffery orbits in viscoelastic fluids. Experiments conducted by \citet{Bartram_1975} in viscoelastic fluids showed that at low shear rates (in comparison to the relaxation time of the fluid) slender bodies tend to see an increase in period of rotation and a drift towards the log-rolling position; however, at higher shear rates, they tend to face the flow direction and stop rotating. \citet{Leal_1975} found similar results, calculated theoretically, for the motion of rod-like particles in second-order fluids. \citet{Brunn1977} found theoretically that the effect of a second-order fluid on ellipsoidal particles in shear flow is as follows: a prolate spheroid drifts to log-rolling, whereas an oblate tumbles around the vorticity axis in direct contradistinction to the effects of weak inertia \cite{Einarsson2015a}. \citet{Gunes2008} carried out experiments for prolate spheroids of moderate aspect ratio in several suspending fluids. They found that elastic effects tend to increase the period of rotation and that the orbits start to drift towards log-rolling. In recent numerical simulations \citet{DAvino2014} have shown that a prolate particle in a viscoelastic fluid, achieves a log rolling orbit at low shear rates while the particle tends to align in the flow direction at high shear rates. 

Much of the work on the dynamics of particles in complex fluids have focused on the effects of viscoelasticity, see for example the recent review by \citet{Shaqfeh_2019} that summarizes prior research and recent advances on the rheology of particle suspensions in viscoelastic fluids. However, many realistic complex fluids tend to exhibit both viscoelasticity and shear-dependent rheology \cite{Brunn1980,Bird1987}. Recently \citet{Datt2018} explored different dynamics of spherical particles in shear-thinning fluids, and observed for example that the rotation rate of a sphere in shear flow is unaffected by shear-thinning rheology, but the impacts of shear rheology on the dynamics of other, (anisotropic) particles have received almost no attention. Recently, \citet{F_rec_2018} investigated the dynamics of  a  two-dimensional ellipsoidal particle in shear flow of a power-law fluid using a finite element simulation. In this proceeding, the authors show a slight reduction in angular velocity that diminishes with particle aspect ratio. \citet{Sadeghy2019} also studied the dynamics of an elliptic particle in a yield-stress fluid using the lattice-Boltzmann method.  In this paper we consider the dynamics of a neutrally buoyant three-dimensional prolate spheroid in a shear flow of a weakly shear-thinning fluid.  The  particle is  small enough that inertial forces are negligible and but big enough that Brownian forces do not play a role. To capture the leading-order effects of shear-thinning rheology on the dynamics, we solve for the motion of the particle for asymptotically weak  shear thinning  using the Carreau model. The question we ask is whether shear-thinning rheology affects the orientational dynamics and to what extent, and in particular, is shear-thinning rheology sufficient to lift the degeneracy of Jeffery orbits in Newtonian fluids. We will show that the orbits are indeed modified but the degeneracy must remain due to the symmetry of the constitutive equations. This paper is organized as follows: we start with the rheology of shear-thinning fluids, then move to the problem of a single prolate spheroid in the linear flow of Newtonian fluid. Finally, we study the deviation of the dynamics due to weak shear thinning.\\

\section{Mathematical model}
\subsection{A spheroid in shear}
\begin{figure}[h!]
	\centering
    	\includegraphics[width=.37\textwidth]{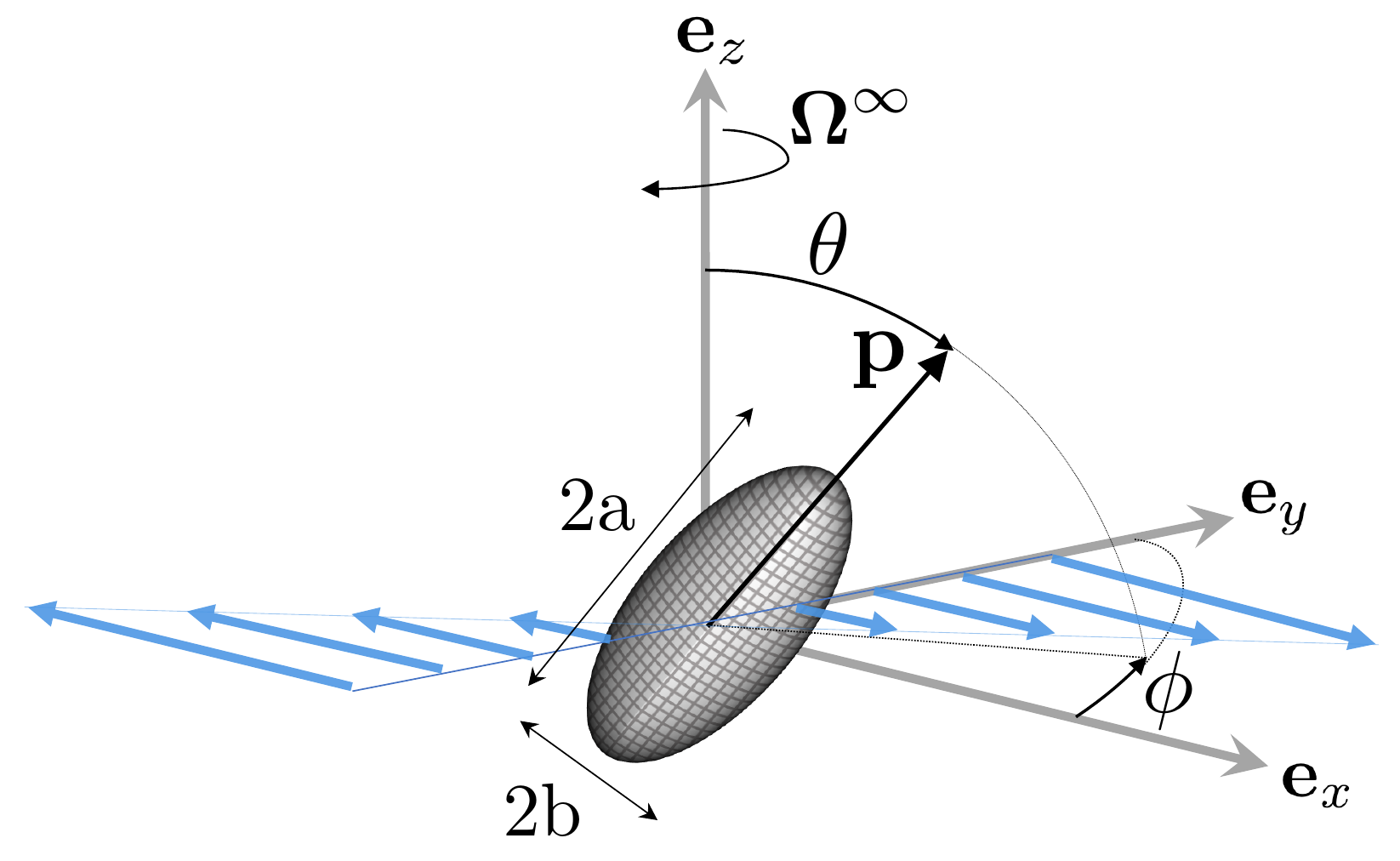}\quad
	\caption{A prolate spheroid in simple shear flow.}
	\label{jc}
\end{figure}

We consider here a prolate spheroid $\fB$ with surface $\partial\fB$, whose axis of symmetry is defined by the unit vector $\bp$ as shown in Fig.~\ref{jc}. The polar angle of the particle, with respect to a fixed lab frame whose origin is at the particle center, is $\theta$ while $\phi$ is the azimuthal angle. The major (minor) axis length is denoted by $a$ $(b)$ and the aspect ratio is $\lambda=a/b=1/{\sqrt{1-e^2}}>1$. The spheroid is immersed in an otherwise linear velocity field defined by  $\bu^{\infty}=\bA^{\infty} \cdot \bx$, whose origin is at the center of the particle to eliminate any constant translation. The velocity gradient tensor $\bnabla\bu^\infty = \bA^\infty$ is constant and may be decomposed in the usual way into symmetric and antisymmetric parts $\bA^\infty = \bE^\infty + \bOmega^\infty\times\bI$ where $\bOmega^\infty$ is the angular velocity of the background flow ($\bI$ is the identity). The flow-shear plane is defined by the basis $\be_x$, and $\be_y$. 

Given a fluid velocity field $\bu$, we define a disturbance velocity field relative to the background as
 \begin{align}
\bu' &=\bu-\bu^{\infty}.
\end{align}
We assume no slip on the surface of the spheroid and that the disturbance flow is zero far from the spheroid, written in terms of the disturbance quantities the boundary conditions are
\begin{align}
\bu' &= (\bOmega-\bOmega^\infty)\times \bx-\bE^{\infty} \cdot \bx  \hspace{1cm}  \bx \in \partial\fB, \\
\bu' &= \bzero \hspace{1cm}  \lb|\bx\rb| \to \infty,
\end{align}
where $\bOmega$ is the angular velocity of the spheroid (the translation velocity $\bU=\bzero$ by construction). The difference in the angular velocity of the spheroid and the undisturbed background flow $\bOmega'=\bOmega-\bOmega^{\infty}$ is sometimes referred to as the slip angular velocity \cite{Einarsson2015b} (although the no-slip condition is indeed satisfied). The evolution of the particle director is
\begin{align}
\dot{\bp} = \bOmega\times\bp.
\end{align}

\subsection{Shear-thinning fluids}
We consider here fluids that shear thin, meaning the viscosity of the fluids, $\eta$, decreases with increasing strain rates, $\bgammad$. To capture this behavior and its effect on the dynamics of a prolate ellipsoid in shear we use the Carreau model \cite{Bird1987} for a generalized Newtonian fluid with deviatoric stress
\begin{align}
\btau =\eta(\dot{\gamma})\dot{\bgamma},
\label{gennewt}
\end{align}
where the functional dependence of the viscosity on the strain-rate,
\begin{align}
\eta(\dot{\gamma})&=\eta_{\infty}+(\eta_{0}-\eta_{\infty}) \big[ 1+\lambda^2_{t} | \dot{\gamma}|^2 \big] ^{\frac{(n-1)}{2}},
\end{align}
is characterized by a zero-shear viscosity $\eta_{0}$, an infinite-shear viscosity $\eta_{\infty}$, a time constant $\lambda_t$, and a power-law index $n<1$. The magnitude of strain-rate is defined $|\dot{\gamma}|=\sqrt{\dot{\bgamma}:\dot{\bgamma}}$. In this study, we explore only the weakly shear-thinning effects on the dynamics of the particle. In this regard, we always assume that $\lambda_t\ll 1/\dot{\gamma}_c$, where we define $\dot{\gamma}_c\equiv\sqrt{2\bE^\infty:\bE^\infty}$ as the characteristic strain rate of the flow defined by the external imposed flow. The deviatoric stress may be conveniently decomposed into Newtonian and non-Newtonian parts, $\btau=\eta_0\bgammad+\btau_{NN}$ where the non-Newtonian part, 
\begin{align}
\btau_{NN}= (\eta(\dot{\gamma})-\eta_0)\dot{\bgamma},
\end{align}
is defined as the amount of  shear thinning  relative to the zero-shear viscosity $\eta_0$.

\subsection{Motion of a particle in a complex fluid}\label{M:C}
In the absence of inertia, the velocity of a particle in a background flow of a fluid of arbitrary rheology may be written as 
\begin{align}
\tU=\tRh_{\tF\tU}^{-1}\cdot\lb[\tF_{ext}+\tF_T+\tF_{NN}\rb],
\end{align}
where $\tU=\left[\bU  \ \bOmega\right]^\top$ is a 6-dimensional vectors containing translational and rotational velocities, likewise $\tF=\left[\bF  \ \bL\right]^\top$ represents both force and torque \cite{Elfring2017}. We consider here a particle that is neutrally buoyant and that no other external force acts on the particle and thus $\tF_{ext}=\bzero$.  

The term
\begin{align}
\tF_T = -\int_{\partial\fB}\bu^\infty\cdot(\bn\cdot\tTh_\tU)\d S,
\end{align}
is the drag force and torque on a particle held fixed in a background flow $\bu^\infty$ of a \textit{Newtonian} fluid of constant viscosity $\eta_0$. In a Newtonian fluid this force, along with the rigid-body resistance tensor,
\begin{equation}
\tRh_{\tF\tU}=
\begin{bmatrix}
\hat{\bR}_{\bF \bU} & \hat{\bR}_{\bF \bOmega}\\
\hat{\bR}_{\bL \bU} & \hat{\bR}_{\bL \bOmega}
\end{bmatrix},
\end{equation}
for the particle in the same Newtonian fluid, would entirely determine the dynamics of a freely moving particle $\tU_0=\tRh_{\tF\tU}^{-1}\cdot\tF_{T}$. 

In a non-Newtonian fluid, there is an extra force/torque on the particle due to the extra deviatoric stress $\btau_{NN}$ in the fluid volume $\fV$ in which the particle is immersed \cite{Leal1980}. This force given by
\begin{align}\label{forcenn_contri}
\tF_{NN} = -\int_{\fV}\btau_{NN}:\tEh_{\tU}\d V.
\end{align}
We may write for simplicity that velocity is composed of a Newtonian part and a non-Newtonian correction $\tU = \tU_0 + \tRh_{\tF\tU}^{-1}\cdot\tF_{NN}$.  A similar approach has also been used for studying the dynamics of active particles in complex fluids \citep{lauga09, natale17, Elfring2017}.

The tensors  $\tEh_{\tU}$ and $\tTh_\tU$,  in addition to $\tRh_{\tF\tU}$, are linear operators that are calculated from the resistance problem for the same particle in a Newtonian fluid with viscosity $\eta_0$. The tensors $\tEh_{\tU}$ and  $\tTh_\tU$ are functions of position in space that map the rigid-body motion $\hat{\tU}$ of a particle to the fluid strain-rate $\bgammadh=2 \tEh_\tU\cdot \tUh$ and stress fields $\bsigmah= \tTh_\tU\cdot \tUh$, respectively.

In our study the translational velocity of the particle is zero, $\bU=\bzero$, and by symmetry the only relevant component of the hydrodynamic resistance is $\hat{\bR}_{\bL \bOmega}$. Because of this the above general expressions simplify considerably so that the angular velocity of the particle in a non-Newtonian fluid is
\begin{align}\label{omega_gen}
\bOmega &=\bOmega_0 - \bRh_{\bL\bOmega}^{-1}\cdot\int_{\fV}\btau_{NN}:\bEh_{\bOmega}\d V,
\end{align}
where the angular velocity in a Newtonian fluid
\begin{align}\label{omega0}
\bOmega_0 = -\bRh_{\bL\bOmega}^{-1}\cdot\int_{\partial\fB}\bu^\infty\cdot(\bn\cdot\bTh_{\bOmega})\d S.
\end{align}
Clearly, the Newtonian dynamics are well known, and to determine the correction we must (only) resolve the integral on the right-hand side of \eqref{omega_gen}. The tensors $\hat{\bR}_{\bL \bOmega}$, $ \hat{\bE}_{\bOmega}$, and $ \hat{\bT}_{\bOmega}$ for prolate spheroids are also well known (see details in appendix \ref{A:E:NN}), but to calculate the integral one must also know the non-Newtonian stress $\btau_{NN}$ in the entire fluid domain and thus requires resolution of the non-Newtonian flow field. To bypass this difficulty we employ a perturbative approach wherein the leading-order contributions to the non-Newtonian stress are determined by the flow-field of the Newtonian solution $\btau_{NN} \sim \btau_{NN}(\bu_0)$.

\subsection{Asymptotic solution}\label{E:N}
First we non-dimensionlize our equations (denoted by *), lengths are scaled by the major axis length $a$ and stresses by $\eta_0 \dot{\gamma}_c$. The dimensionless non-Newtonian stress is thus
\begin{align}
\btau_{NN}^*= -(1-\beta)\Big(1-[1+Cu^2|\dot{\gamma}^*|^2]^{(n-1)/2}\Big)\dot{\bgamma}^*.
\end{align}
The Carreau number $Cu=\dot{\gamma}_c\lambda_t$ is the ratio of the characteristic strain rate $\dot{\gamma}_c$ to the crossover strain-rate $1/\lambda_t$, while the viscosity ratio is $\beta=\eta_{\infty}/\eta_{0}$.  We note that when $Cu=0$ or $\beta=1$ the fluid is Newtonian. For weak deviations from Newtonian behaviour one may take as a small parameter $Cu^2$ or $1-\beta$ \cite{Datt2015}, here we choose $Cu^2$ to explore the first effects of shear-thinning as this leads to a much more analytically tractable expression. Thus, flow quantities are expanded  in regular perturbation series in powers of $Cu^2$, $\bu^*=\bu^*_0+Cu^2 \bu^*_1+\mathcal{O}(Cu^4)$, and $\btau^*=\btau^*_0+Cu^2 \btau^*_1+\mathcal{O}(Cu^4)$ where $\bu^*$, and $\btau^*$ are the dimensionless velocity field, and deviatoric stress fields respectively. In this way, the non-Newtonian deviatoric stress
\begin{align}
\label{tau:NNN} 
\btau_{NN}^*&=Cu^2 \btau_{NN,1}^* +\mathcal{O}(Cu^4)\nonumber\\
&=  -\frac{1}{2}Cu^2(1-\beta)(1-n)|\dot{\gamma}^*_0|^2 \dot{\bgamma}^*_0+\mathcal{O}(Cu^4).
\end{align}

Writing similarly for the orientational dynamics of the spheroid $\bOmega^* = \bOmega_0^* + Cu^2\bOmega_1^* + \fO(Cu^4)$ we find, by way of equations \eqref{omega_gen} and \eqref{tau:NNN}, that
\begin{align}\label{omega1}
\bOmega_1^* &= - \bRh_{\bL\bOmega}^{*-1}\cdot\int_{\fV}\btau_{NN,1}^*:\bEh_{\bOmega}^*\d V^*\nonumber\\
&=\frac{1}{2}(1-\beta)(1-n)\bRh_{\bL\bOmega}^{*-1}\cdot\int_{\fV}|\dot{\gamma}^*_0|^2 \dot{\bgamma}^*_0:\bEh_{\bOmega}^*\d V^*.
\end{align}
We see that the change in the orientational dynamics depends only on the Newtonian velocity field $\bu_0^*$ to leading order in $Cu$. The main mathematical task of this work is to calculate and integrate the tensor $|\dot{\gamma}^*_0|^2 \dot{\bgamma}^*_0:\bEh_{\bOmega}^*$ over the entire fluid domain $\fV$. This is a rather involved integral and therefore we want to use as amenable a representation of the Newtonian solution as possible. To this end, we use a solution based on an expansion in spheroidal multipoles taken directly from \citet{Einarsson2015b}. For completeness we repeat details of that solution in the appendix of this work.

For simplicity we now drop the *'s and use only dimensionless variables from this point on unless explicitly stated otherwise.
 \section{Results}
\subsection{Jeffery Orbits} 
The solution to the zeroth order (Newtonian) problem yields the classical Jeffery orbits of an ellipsoidal particle in shear flow. The solution obtained by \citet{Jeffery1922} involved solving for the disturbance flow field, but it can also be found by directly integrating \eqref{omega0}. The angular velocity of a prolate ellipsoid in a Newtonian fluid is
\begin{align}
\bOmega_0 =\bOmega^{\infty}+\Lambda\, \bp \times \bE^{\infty} \cdot \bp,
\end{align}
where $\Lambda=\frac{\lambda^2-1}{\lambda^2+1}$, while the evolution of the director is
\begin{align}
\bpd = \bOmega_0\times\bp = \bOmega^{\infty}\times\bp+\Lambda\, (\bI-\bp\bp)\cdot\bE^\infty\cdot\bp.
\end{align}
Jeffery's results show that, unlike a point particle or a sphere, a prolate spheroid rotates not only with the (constant) local angular velocity of the flow but, given that object is elongated and unevenly samples the velocity field about its center, has a rotational component that depends on the orientation and aspect ratio of the spheroid. 

\begin{figure}
	\centering
	\includegraphics[trim={0cm 0 0cm 0},clip,width=.375\textwidth]{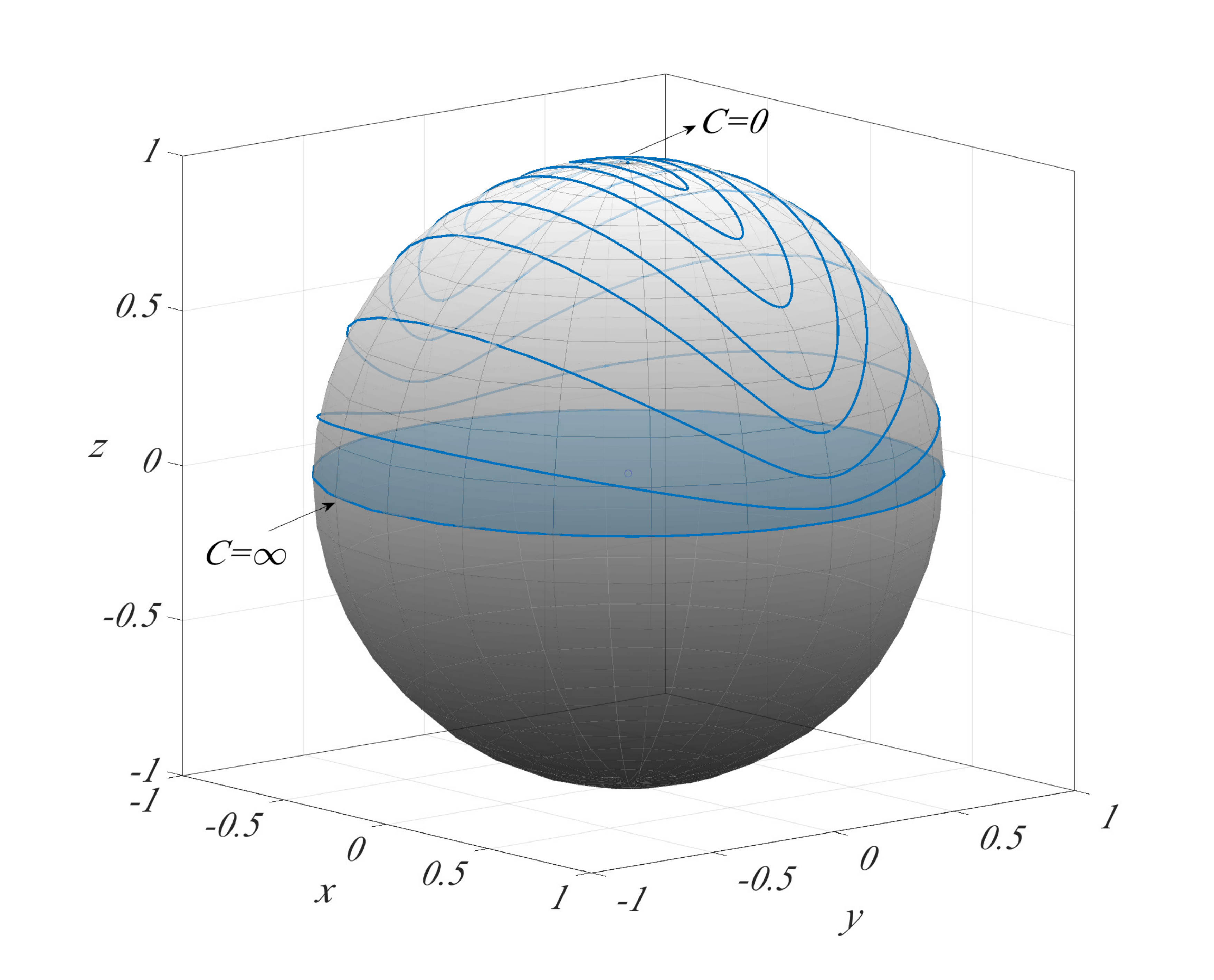}
	\caption{ Trajectories in the orientation of a prolate spheroid with aspect ratio $\lambda=5$ in a linear shear flow of Newtonian fluid, for various different initial positions. }
	\label{jefferyorbits}
\end{figure}

Assuming a background flow field (as we do throughout this paper)
\begin{align}
\bu^\infty = y\be_x
\end{align} 
and $\phi=0$ at $t=0$, leads to Jeffery orbits of the form
\begin{align}
&\tan\phi=\lambda \tan\left( \frac{\lambda t}{1+\lambda^2}\right),\nonumber\\
&\tan\theta=\frac{C \lambda}{\sqrt{\text{sin}^2\phi+\lambda^2\text{cos}^2\phi}},
\label{orbits} 
\end{align}
where $C$ is a constant of integration and the axis of revolution rotates in one of infinitely many possible periodic orbits depending on the value of $C$. Fig.~\ref{jefferyorbits} shows Jeffery orbits for different values of $C$ on a unit sphere. The Jeffery orbit on the equator of the sphere, $C \to \infty$, is called the tumbling orbit because the vector $\bp$ tumbles in the flow-shear plane. The orbit at the pole of the sphere, $C=0$, where $\bp$ is aligned with the vorticity direction, is called log-rolling. The period $T_0=2 \pi (\lambda^2+1)/\lambda$ is constant for particles of same aspect ratio and does not depend on the initial orientation of the particle (in dimensional terms the period scales with $1/\gammad_c$). Note that the angular velocity of the particle is not constant in time but that the particle slows down when $\bp$ tends to the flow direction \cite{DAvino2015}.

\subsection{Dynamics of a prolate spheroid in shear flow of a shear-thinning fluid}\label{Num:results}
The zeroth order (Newtonian) solution of the flow field is linear, and thus we must be able to write $\bgammad_0 = 
\bM:\bE^\infty$ where $\bM$ is a fourth order tensor that depends on the orientation of the particle alone. Substituting into \eqref{omega1} we obtain for each component of $\bOmega_1$
\begin{align}
\Omega_{1j}&=\frac{ (1-\beta)(1-n)}{2} \hat{R}_{L \Omega, jt}^{-1} E^{\infty}_{ls}E^{\infty}_{pq}E^{\infty}_{gf}
\int_\fV  M_{imsl}M_{mipq} M_{uvfg}  \hat{E}_{\Omega,vut} \d V,
\end{align}
where repeated indices are summed. Although $\bM$ is constructed from a known Newtonian solution the details are quite complicated (as shown in appendix ). This tensor product contains hundreds of terms and analytical evaluation of the integral proves more or less intractable and so this integral is performed numerically.  After calculation of all the tensors, a trapezoidal rule in spheroidal coordinates is used for the integration with singular terms evaluated analytically.

Upon resolution of $\bOmega_1$ we calculate the periodic orbits of $\bp$ in a shear-thinning fluid to leading order in $Cu$, namely we integrate
\begin{align}
\dot{\bp} = (\bOmega_0+Cu^2\bOmega_1)\times\bp
\end{align}
forward in time. An \textit{RK4} method is used for the time derivative of the particle's orientation, and the orientation is expressed by angles $\theta$ and $\phi$ rather than vector $\bp$ to ensure unity of its magnitude.

The dynamics of the spheroid can be divided into a rotation around the vorticity axis, $\phi$ (spinning), and rotation about the velocity axis, $\theta$ (oscillating). While the period in a Newtonian fluid, $T_0$, is determined only by the particle shape and shear-rate, the period in a non-Newtonian fluid, $T$, is highly dependent on initial position of the particle and $Cu$ number in shear-thinning fluids.

\begin{figure}
  \centering
  \subcaptionbox{}[.23\textwidth][c]{%
    \includegraphics[trim={7.5cm 3cm 0cm 4cm},clip,width=.36\textwidth]{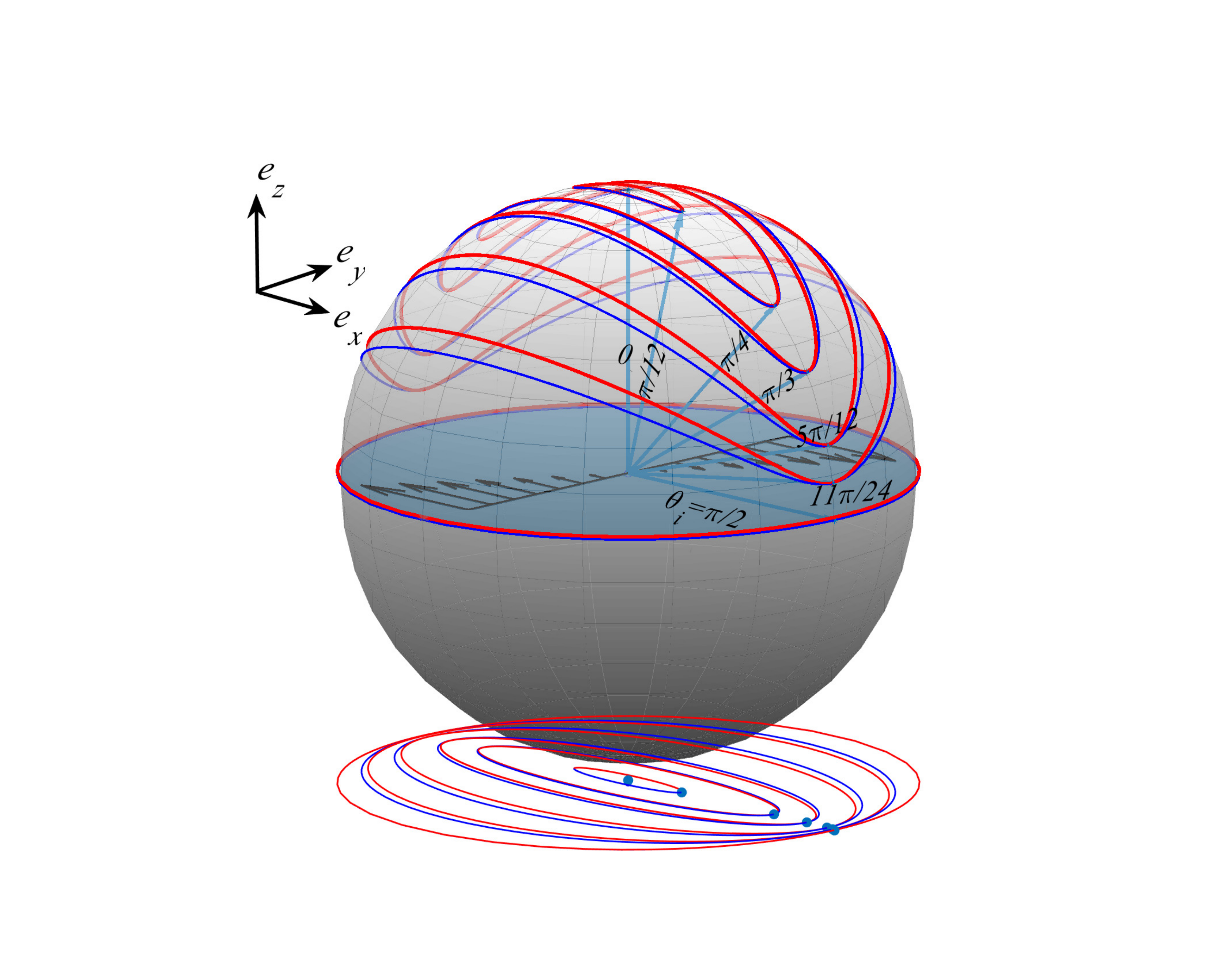}}\quad
  \subcaptionbox{}[.23\textwidth][c]{%
    \includegraphics[trim={7.5cm 3cm 0cm 4cm},clip,width=.36\textwidth]{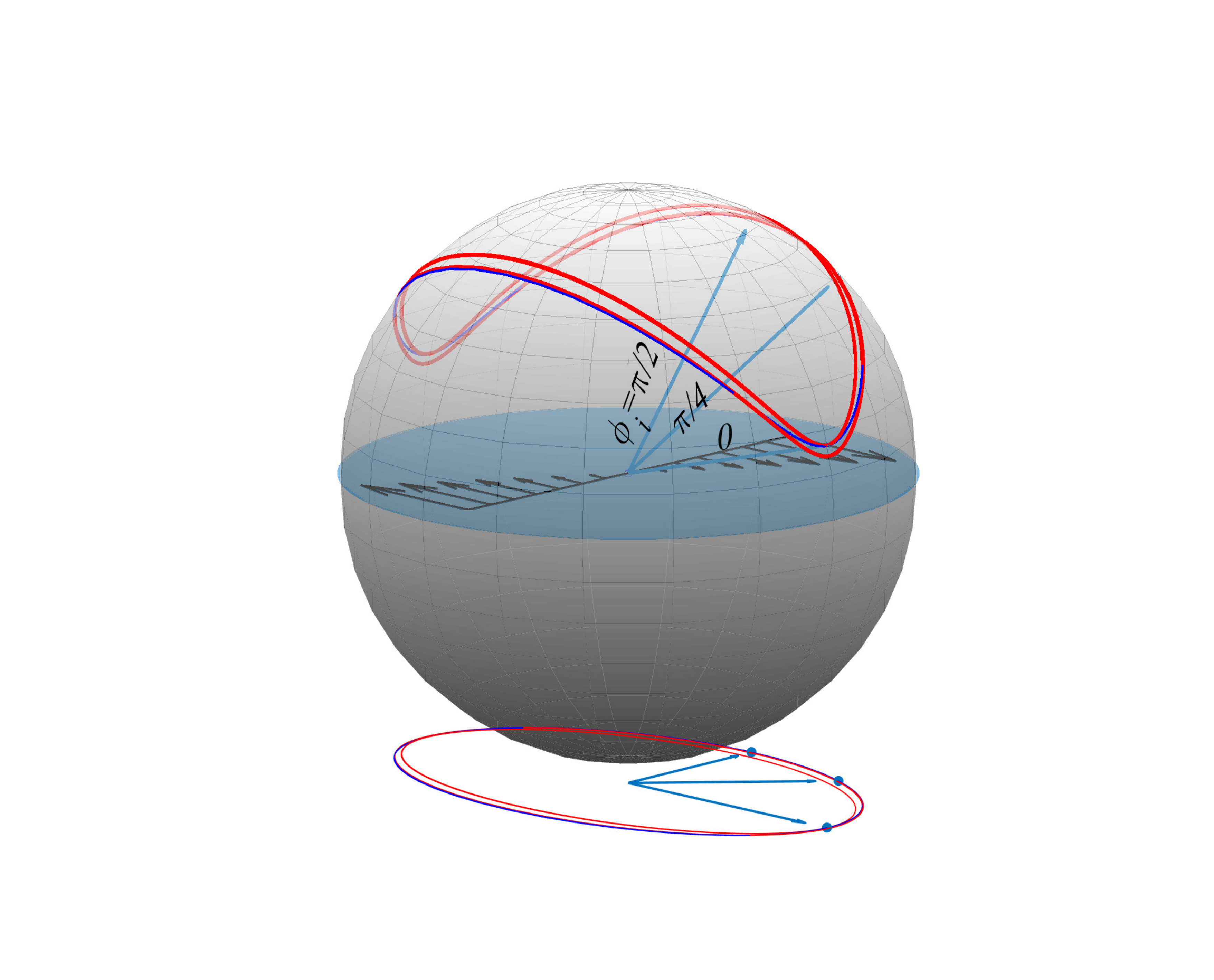}}\quad
   \subcaptionbox{}[.23\textwidth][c]{%
    \includegraphics[trim={7.5cm 3cm 0cm 4cm},clip,width=.36\textwidth]{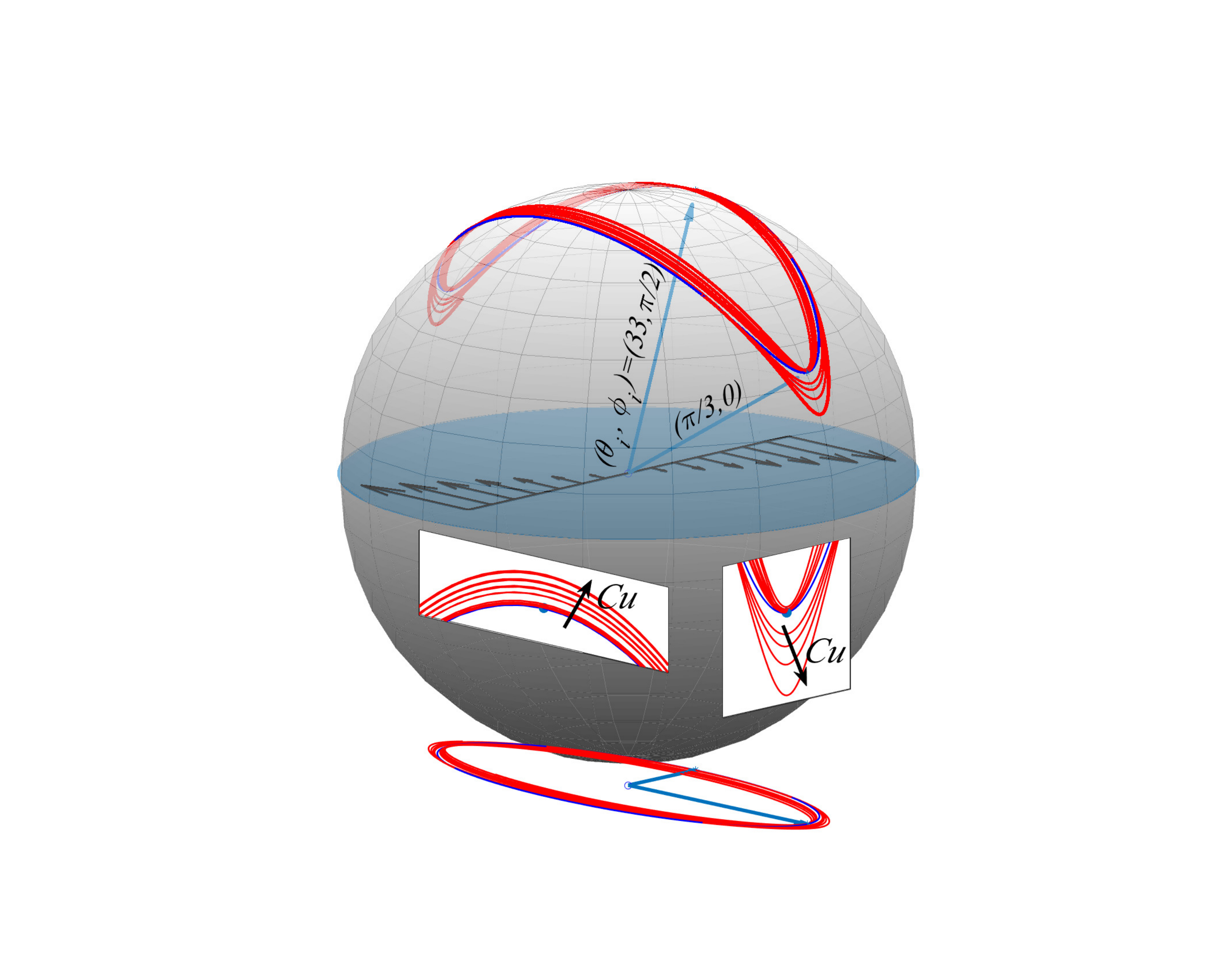}}\quad
      \subcaptionbox{}[.23\textwidth][c]{%
    \includegraphics[trim={7.5cm 3cm 0cm 4cm},clip,width=.36\textwidth]{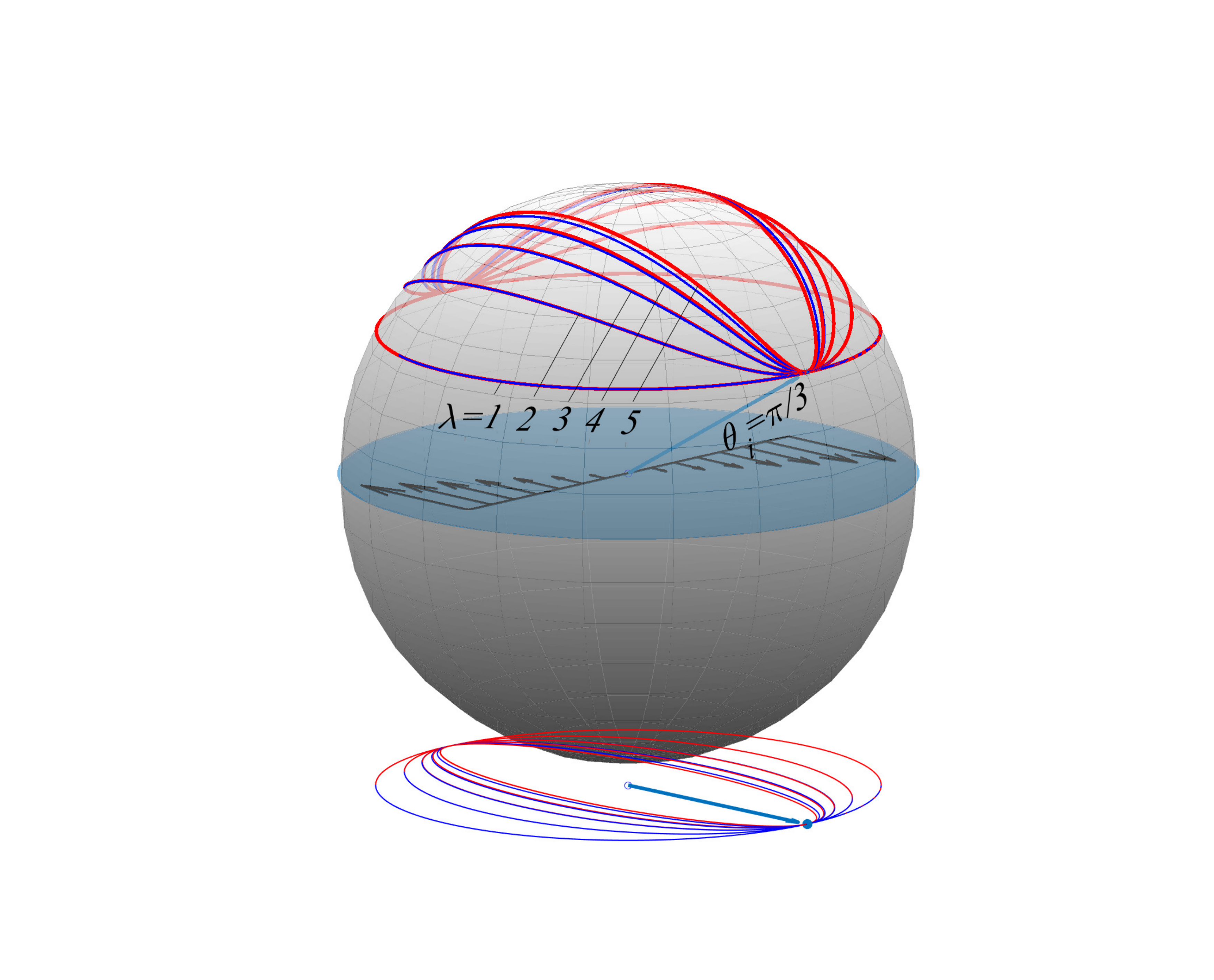}}\quad
  \caption{Modified orbits (red lines) in the presence of shear thinning and Newtonian orbits (blue lines) for
  a) different initial conditions of $\theta_i=0,\pi/12,\pi/4,\pi/3,5\pi/12,11\pi/24,\pi/2$ and $\phi_i=0$, 
  b) different initial conditions of $\phi_i=\pi/2, \pi/4, 0$ on a specific Jeffery orbit passing $(\theta_i,\phi_i)=(5\pi/12,0)$,
  c) different Carreau numbers of $Cu=0, 0.1, 0.12, 0.14, 0.16$ for two different initial conditions of $(\theta_i,\phi_i)=(\pi/3,0), (0.33, \pi/2)$ belonging to a Jeffery orbit, and
  d) different values of aspect ratios $\lambda=1,2,3,4,5$ for $(\theta_i,\phi_i)=(\pi/3,0)$.
 Calculations are carried out for $Cu=0.1$, and $\lambda=5$ unless otherwise stated.}
  \label{result:orbit}
\end{figure}

In Fig.~\ref{result:orbit} we show Jeffery orbits modified by shear thinning for different situations: a) different initial positions $\theta_i$ (for $\phi_i=0$), b) different initial positions on a specific Newtonian Jeffery orbit passing $(\theta_i,\phi_i)=(5\pi/12,0)$ c) different values of $Cu$, and finally d) different aspect ratios $\lambda$. In these results and all that follow we take values of $\beta=0.5$, and $n=0.5$. We also take $Cu=0.1$, and $\lambda=5$ unless otherwise stated. The first thing to notice in the figures is that shear-thinning rheology does not lift the degeneracy of the Jeffery orbits observed in Newtonian fluids, unlike the effects of fluid elasticity or inertia. There are still infinitely many modified `Jeffery' orbits, selected by the initial condition, that repeat periodically for all time. Indeed this is somewhat expected given that the generalized Newtonian fluid constitutive equation \eqref{gennewt}, maintains the symmetries of the Stokes equations and so one should not expect a symmetry-breaking drift of the orbits in time. In general we observe that  shear thinning  tends to narrow the orbits in much the same way that an elongation of the particle aspect ratio does (compare Fig.~\ref{result:orbit}a), c) with d).  The change in a particular Newtonian Jeffery orbit due to shear thinning depends on the initial position as shown in Fig.~ \ref{result:orbit}b), because of course the orbits are not continuously overlapping.

\begin{figure}
  \centering
  \subcaptionbox{}[.3\textwidth][c]{%
    \includegraphics[trim={1cm 1 1cm 1},clip,width=.35\textwidth]{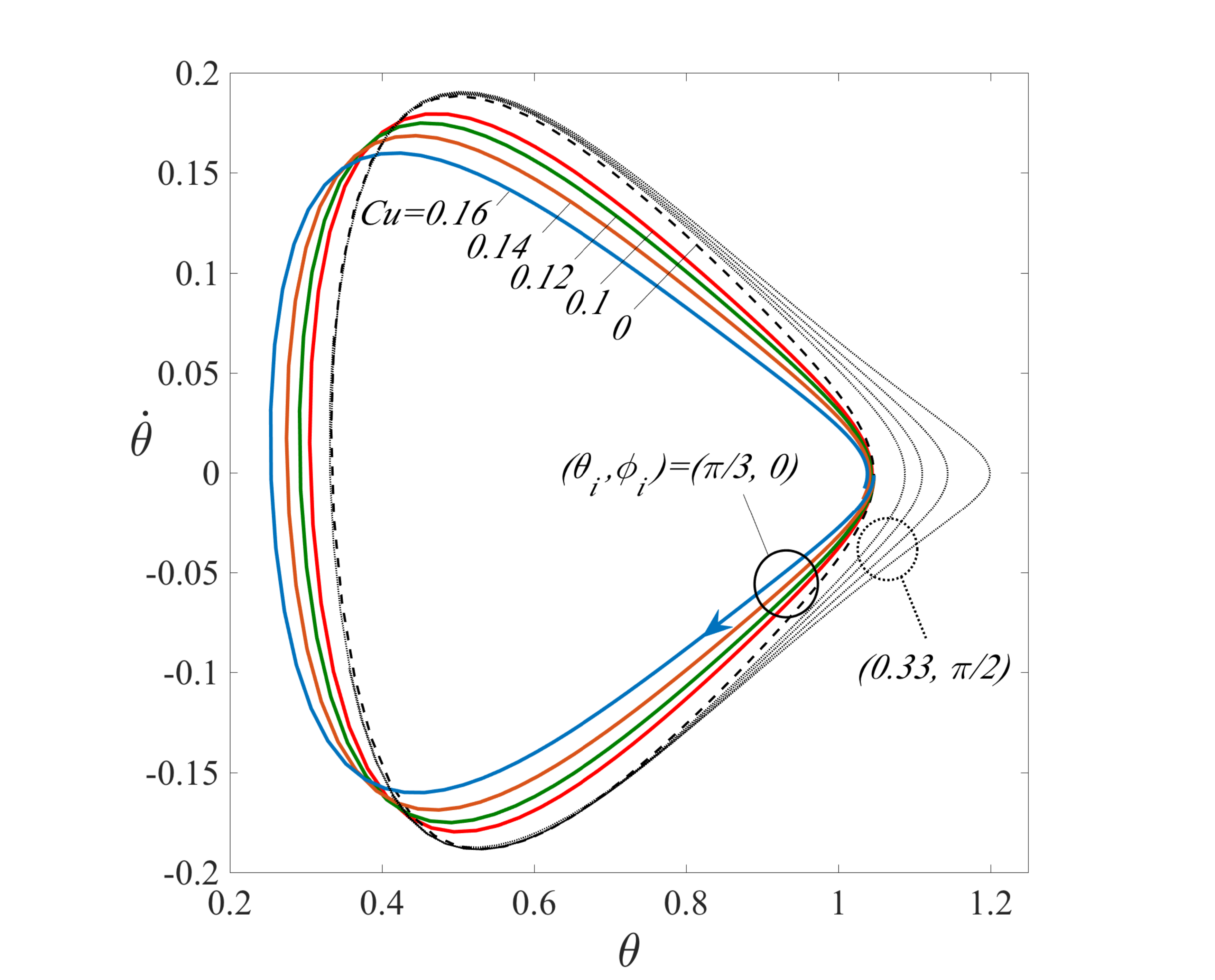}}\quad
  \subcaptionbox{}[.3\textwidth][c]{%
    \includegraphics[trim={1cm 1 1cm 1},clip,width=.35\textwidth]{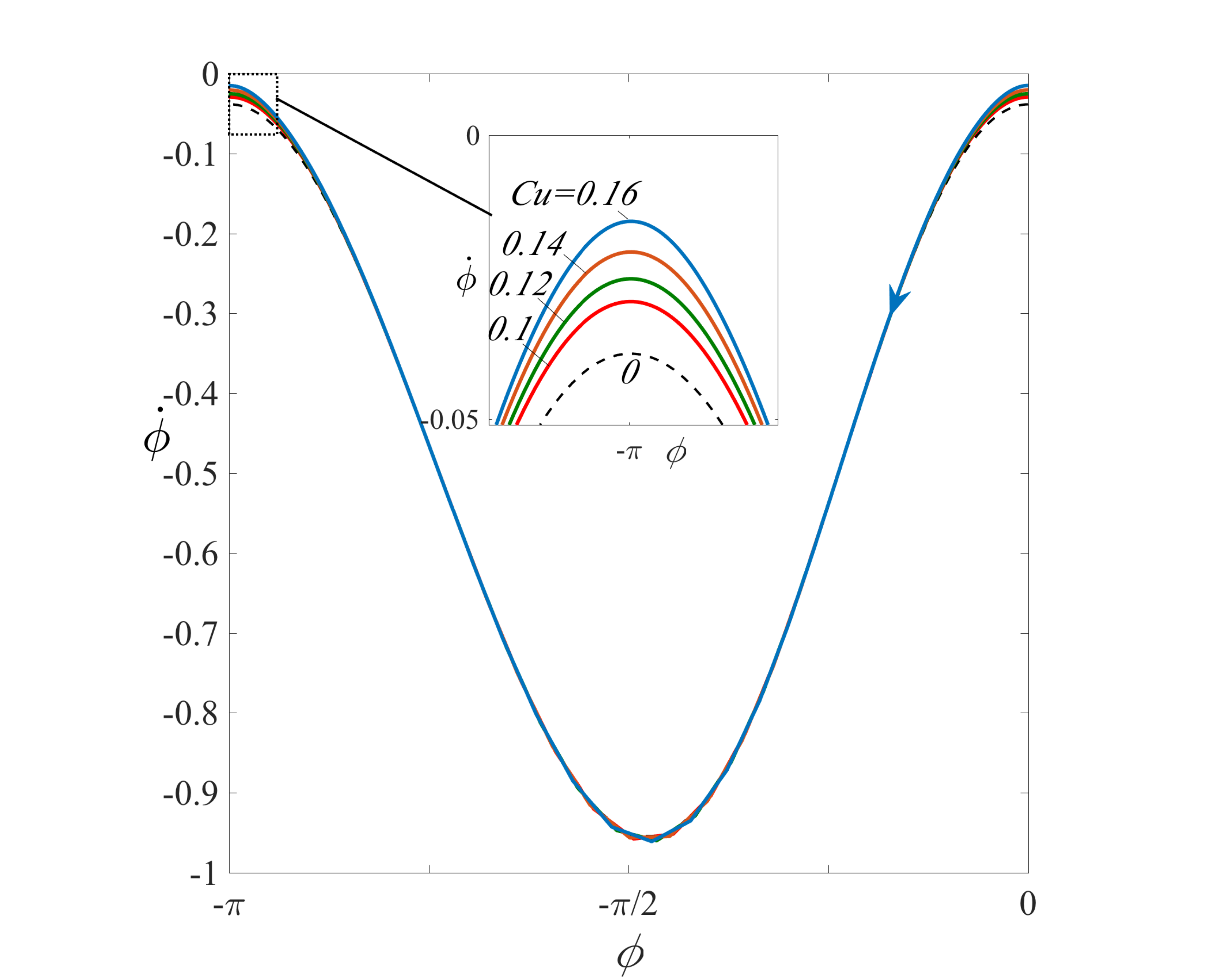}}\quad
      \subcaptionbox{}[.3\textwidth][c]{%
    \includegraphics[trim={1cm 1 1cm 1},clip,width=.35\textwidth]{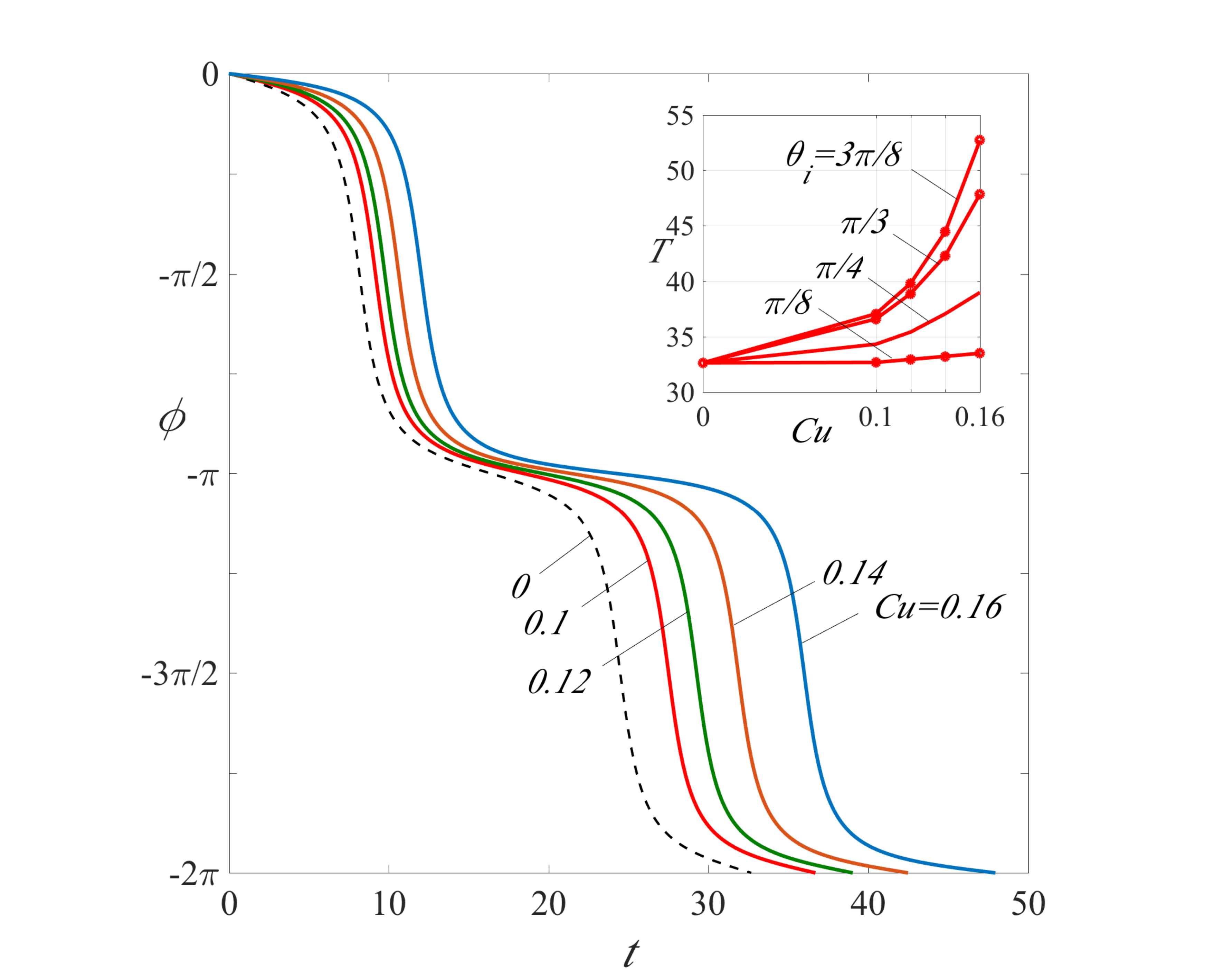}}\quad
  \caption{Phase portraits of a) $\theta$ and b) $\phi$, while c) shows $\phi(t)$ over one full period (with the period, $T$, shown in the inset). Various different $Cu$ numbers are shown while dashed-lines indicate the Newtonian case. Results are for  $(\theta_i, \phi_i)=(\pi/3,0)$, and $\lambda=5$, while the dotted-lines in a) show $(\theta_i, \phi_i)=(0.33,\pi/2)$.}
\label{result:1}
\end{figure}

As the aspect ratio of the particle increases, the particle spends a larger amount of time aligned with the flow as the torque due to the applied background flow is diminished in comparison to the hydrodynamic resistance to rotation, as dictated by \eqref{omega0}, and thus the period of rotation increases. Much  of  the same thing happens with  shear thinning , where the torque is reduced due  to  changes in the viscosity. These changes in the dynamics are illustrated in Fig.~\ref{result:1}a) and \ref{result:1}b). In particular, note in Fig.~\ref{result:1}b)  that the effects of  shear thinning  on the particle spin ($\phid$) are only apparent when the particle is aligned with the flow, $\phi \approx n\pi$, where shear dominates and hence the particle is slowed further due to a reduction of the viscosity. When the (slender) particle is aligned with the velocity  gradient, changes in viscosity are less relevant, as the particle is essentially pushed around its orbit . Similar results were given for two-dimensional particles in \citet{F_rec_2018}. We note that while changes in the absolute value of the the rotation rate might be small, the changes in the period of rotation can be dramatic if the angular velocity is close to zero as shown in Fig.~\ref{result:1}c); however, while the instantaneous velocity of the particle may be accurate in either the Newtonian case, or the correction we calculate here, neglected effects such as inertia, particle eccentricity and idealizations used in Carreau model, will cumulatively affect the orbit of particle integrated over time.

\begin{figure}
  \centering
  \subcaptionbox{}[.3\textwidth][c]{%
    \includegraphics[trim={1cm 1 1cm 1},clip,width=.35\textwidth]{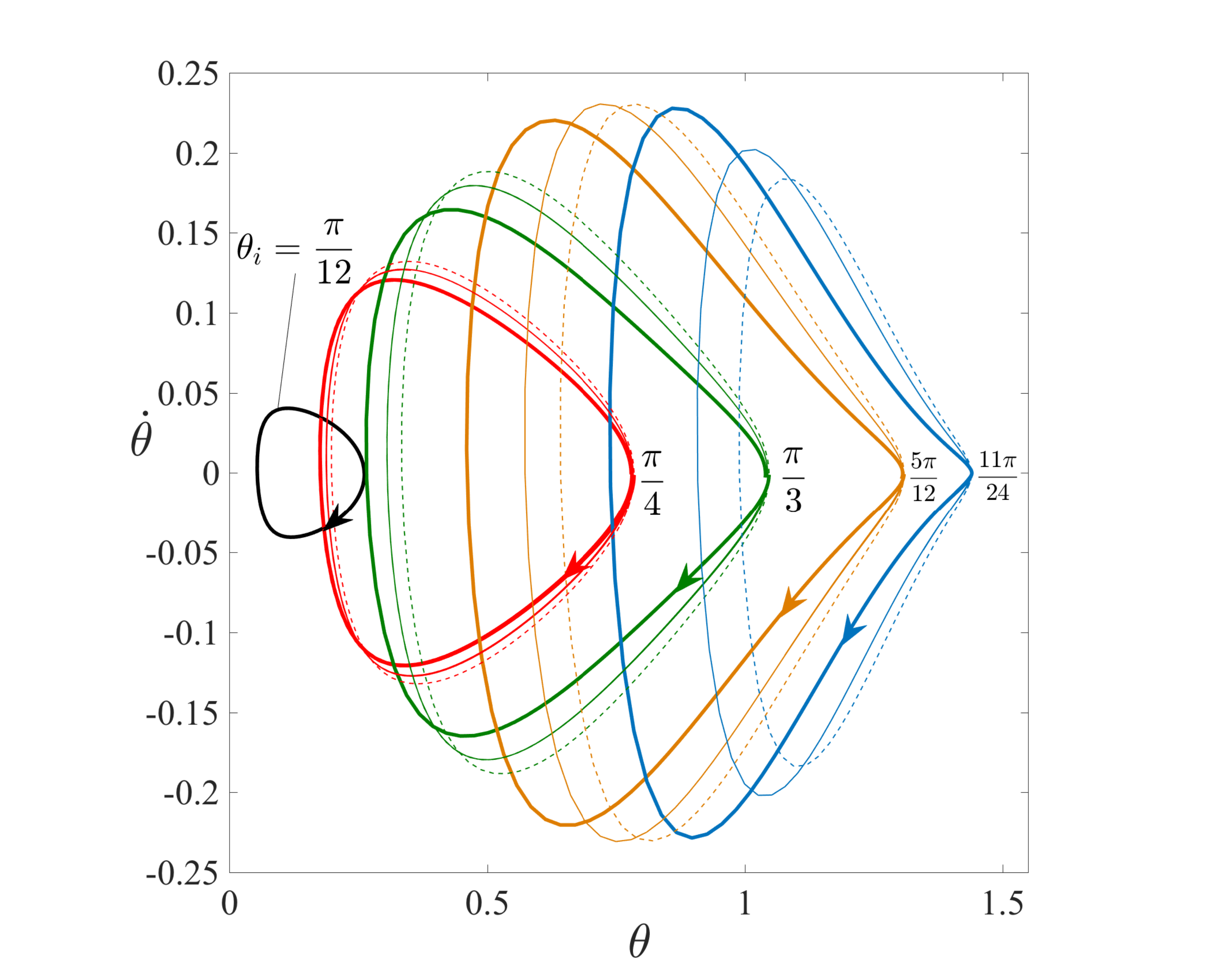}}\quad
  \subcaptionbox{}[.3\textwidth][c]{%
    \includegraphics[trim={1cm 1 1cm 1},clip,width=.35\textwidth]{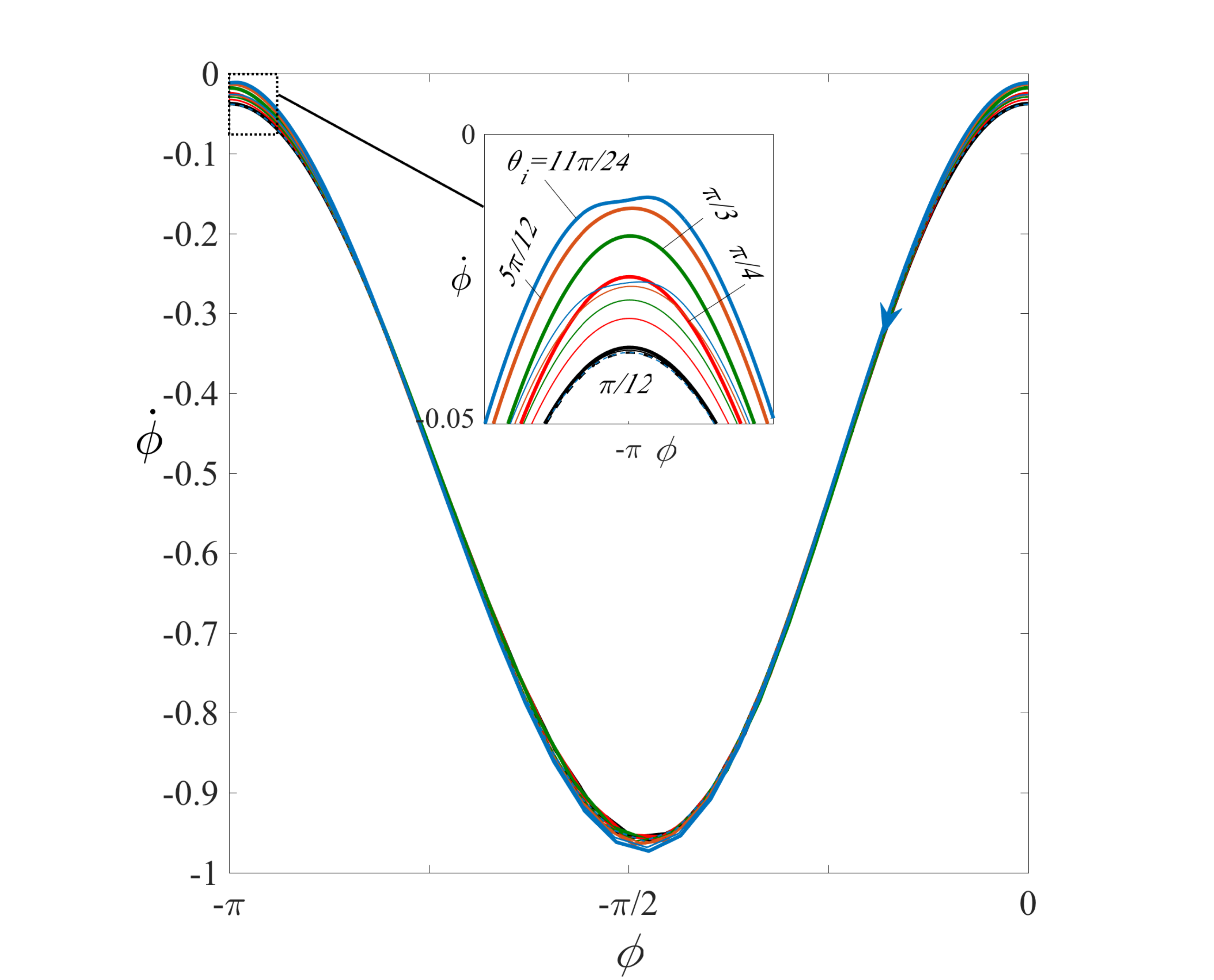}}\quad
      \subcaptionbox{}[.3\textwidth][c]{%
    \includegraphics[trim={1cm 1 1cm 1},clip,width=.35\textwidth]{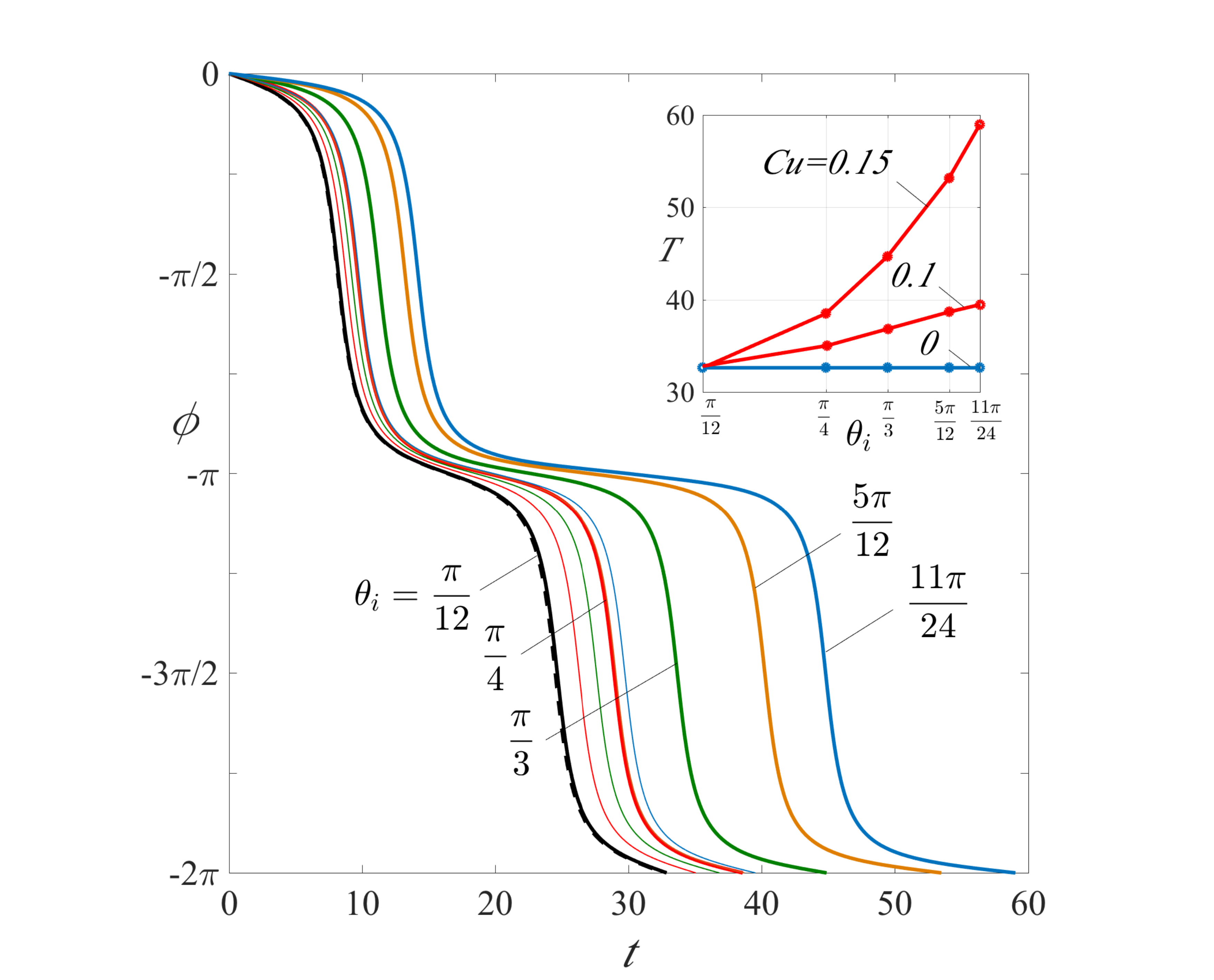}}\quad
  \caption{Phase portraits of a) $\theta$ and b) $\phi$, while c) shows $\phi(t)$ over one full period (with the period, $T$, shown in the inset), for different initial conditions of $\theta_i=\pi/12,\pi/4,\pi/3,5\pi/12,11\pi/24$. Thick, and thin lines correspond to $Cu=0.15$, and $Cu=0.1$ respectively, while dashed-lines indicate the Newtonian case. Results are for $\lambda=5$.}
\label{result:2}
\end{figure}

As shown in Fig.~\ref{result:2}, the effect of  shear thinning  varies quantitatively depending on the particular orbit but the qualitative picture is similar. In particular, the period of each orbit depends on the initial position, unlike in a Newtonian fluid, in a shear-thinning fluid each orbit has a different period.

\begin{figure}
  \centering
  \subcaptionbox{}[.3\textwidth][c]{%
    \includegraphics[trim={1cm 1 1cm 1},clip,width=.35\textwidth]{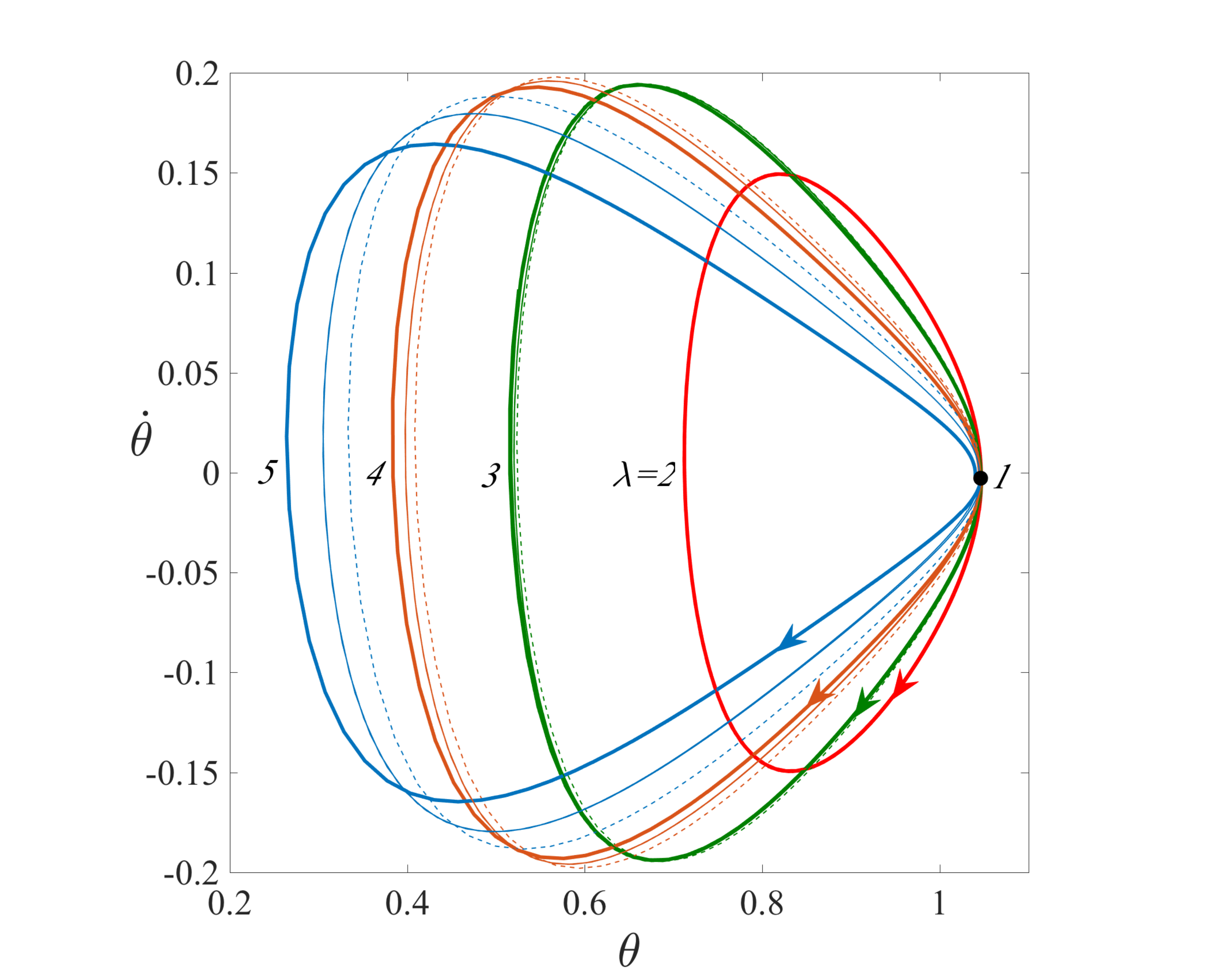}}\quad
  \subcaptionbox{}[.3\textwidth][c]{%
    \includegraphics[trim={1cm 1 1cm 1},clip,width=.35\textwidth]{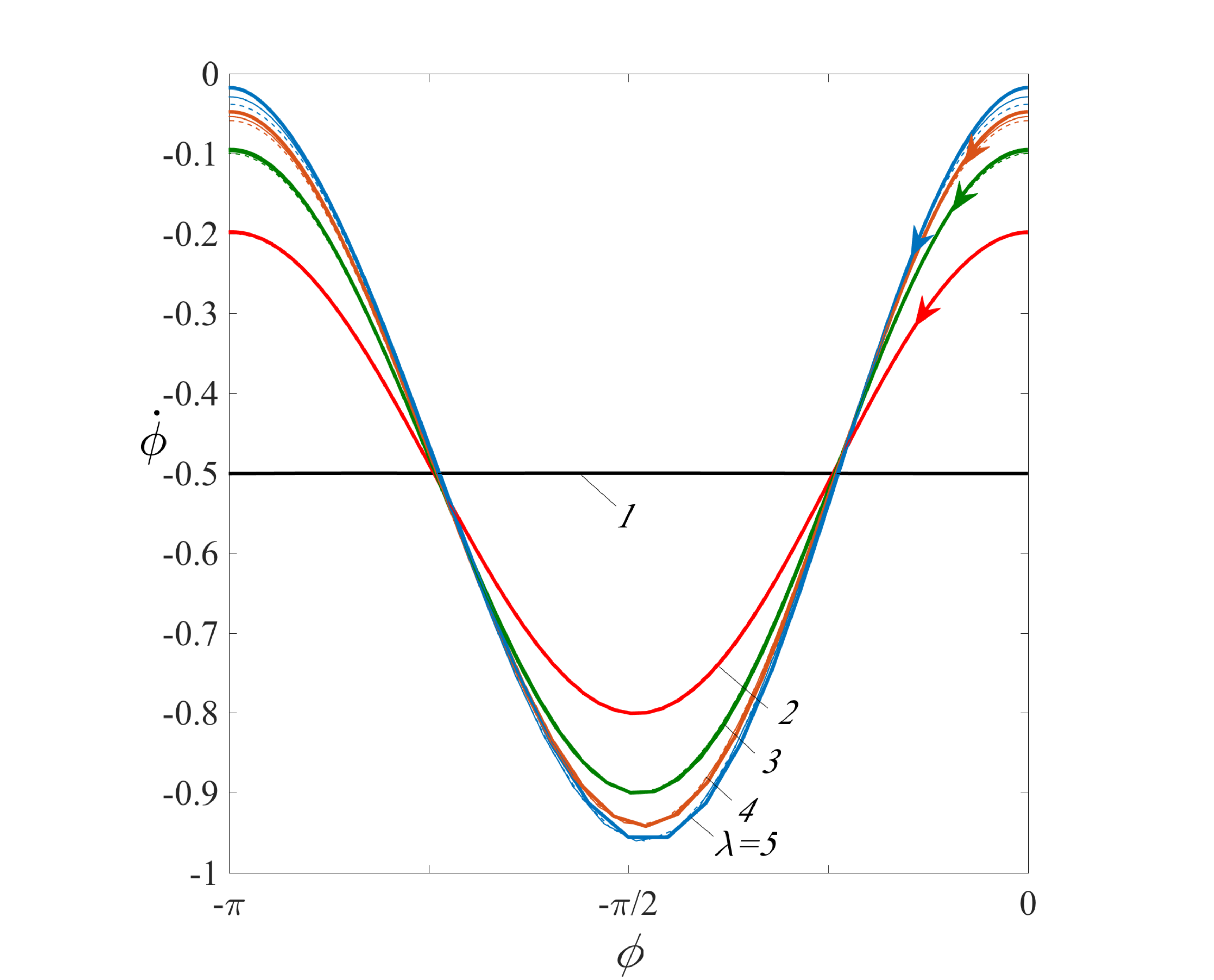}}\quad
      \subcaptionbox{}[.3\textwidth][c]{%
    \includegraphics[trim={1cm 1 1cm 1},clip,width=.35\textwidth]{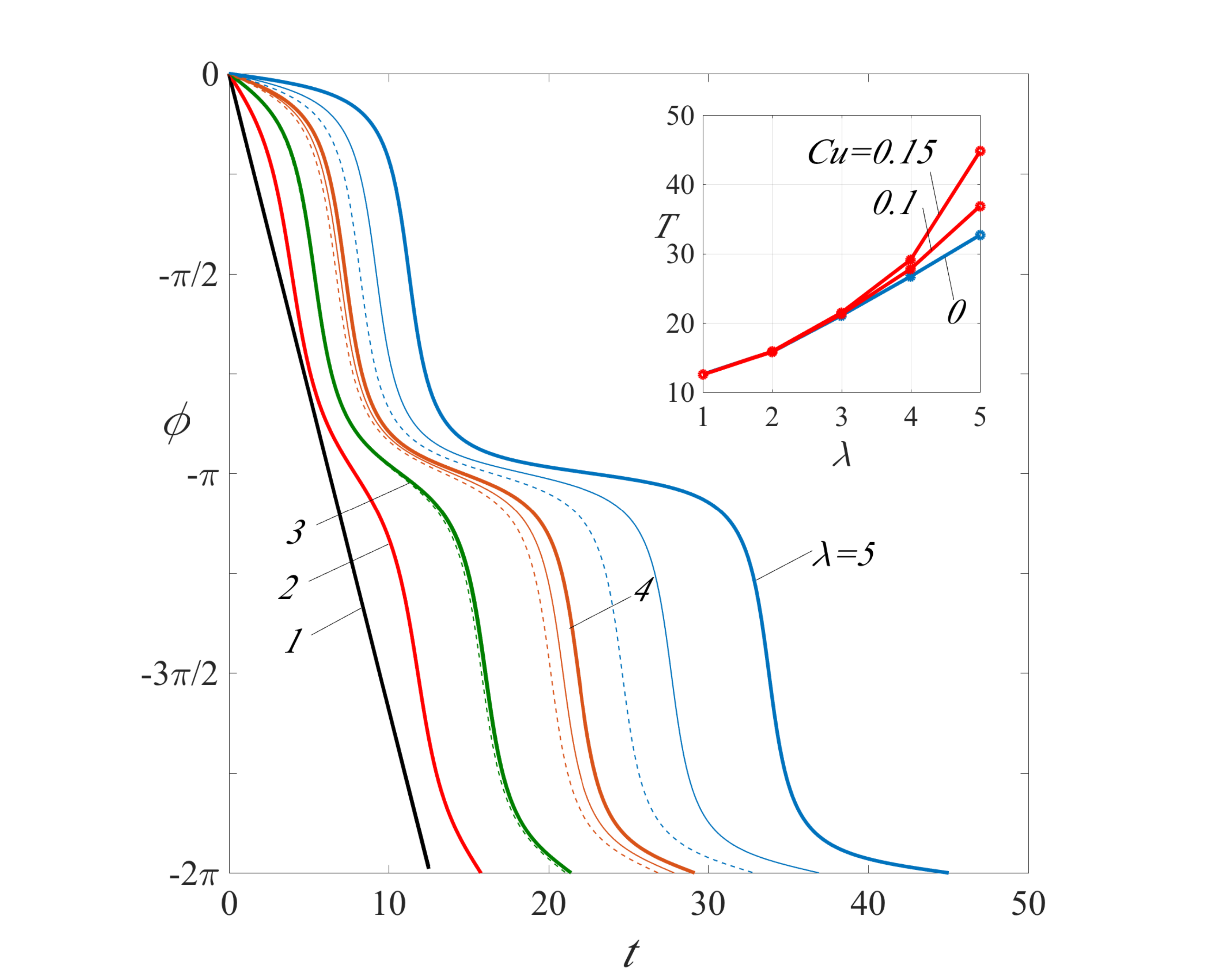}}\quad
  \caption{Phase portraits of a) $\theta$ and b) $\phi$, while c) shows $\phi(t)$ over one full period (with the period, $T$, shown in the inset), for different aspect ratios of $\lambda=1,2,3,4,5$. Thick, and thin lines correspond to $Cu=0.15$, and $Cu=0.1$ respectively, while dashed-lines indicate the Newtonian case.  Results are carried out for $(\theta_i,\phi_i)=(\pi/3,0)$.}
  \label{result:3}
\end{figure}

Different orbits for different aspect ratios, $\lambda$, are shown in Fig.~\ref{result:3}. In case of a sphere, $\lambda=1$, the period doesn't change at all. In other words,   the spherical particle is unaffected by shear thinning,  as discussed in \citet{Datt2018}. For higher aspect ratios, the orbit slows when the particle is aligned with the flow as shown in Fig.~\ref{result:3}b), and the period $T$ thereby increases, as shown in Fig.~\ref{result:3}c). We find that shear-thinning rheology \textit{exacerbates} this effect, more substantially increasing the period for larger $\lambda$ as the shear stresses, which are needed to rotate a long slender particle aligned with the flow, are reduced by shear thinning.

%
\section{Conclusion}\label{Conclusion}
In this work, we investigated the orientational dynamics of a prolate spheroid immersed in a background shear flow of a shear-thinning Carreau fluid. An equation of motion for the rotation of the prolate particle was derived for asymptotically weak  shear thinning  using a regular perturbation expansion in the Carreau number and then integrated numerically. We found that shear-thinning rheology does not lift the degeneracy of the Jeffery orbits observed in Newtonian fluid. In shear-thinning fluids there are still infinitely many orbits that repeat periodically for all time, each selected by the initial condition. However, the instantaneous rotation rate and trajectories of the orbits are modified. Qualitatively,  shear thinning  has a similar effect as elongating a particle in a Newtonian fluid:  shear thinning  tends to increase the period of rotation as the particle slows down more when aligned with the flow due to a reduction of shear stress. Unlike for Jeffery orbits in Newtonian fluids, in shear-thinning fluids the period of the orbits does depend on the specific trajectory (or initial orientation of the particle) because the effect of  shear thinning  varies depending on the orientation of the particle.  The results presented in this work can serve as a base for further investigation into the rheology of anisotropic particles suspended in shear-thinning fluids.

\acknowledgements
We thank Giovanniantonio Natale for discussions that motivated this work. Funding from the Natural Sciences and Engineering Research Council of Canada (NSERC) is gratefully acknowledged.

\bibliography{reference}
%

\newpage
\appendix
\section{Spheroidal multipoles}\label{A:E:N}
Here we give the solution to the Stokes equations for a spheroid with the aspect ratio of $\lambda$ in a shear flow, following \citet{Einarsson2015b}, in terms of a finite multipole expansion \cite{Chwang1975a}.\\

To this end the Green's function, $\bG$, of the Stokes equations and derivatives will be utilized, these are, in component form
 \begin{align}
G_{ij}&=\frac{\delta_{ij}}{r} +\frac{x_ix_j}{r^3}, \hspace{1cm} \text{Stokeslet}, \\
G^d_{ijk}&=G_{ij,k}=\frac{\delta_{jk}x_i+\delta_{ik}x_j-\delta_{ij}x_k}{r^3} -3\frac{x_ix_j}{r^5}, \hspace{1cm} \text{dipole},\\
G^D_{ij}&=G_{ij,ll}=2\frac{\delta_{ij}}{r^3} -6\frac{x_ix_j}{r^5},\hspace{1cm} \text{potential doublet},\\
G^R_{ijk}&=\frac{1}{2} (G_{ij,k}-G_{ik,j})=\frac{\delta_{ik}x_j-\delta_{ij}x_k}{r^3}, \hspace{1cm} \text{rotlet},\label{Dgreen}\\
G^{S}_{ijk}&=\frac{1}{2} (G_{ij,k}+G_{ik,j})=\frac{\delta_{kj}x_i}{r^3}-3\frac{x_ix_jx_k}{r^5}, \hspace{1cm} \text{stresslet},\\
G^Q_{ijk}&=G_{ij,llk}=-6\frac{\delta_{jk}x_i+\delta_{ik}x_j+\delta_{ij}x_k}{r^5} +30 \frac{x_ix_jx_k}{r^7},\hspace{1cm} \text{potential quadrupole}.
\end{align}

Representation of the flow around a spheroidal particle requires a weighted distribution of the above multipoles. Spheroidal multipoles are found by employing a line distribution of Stokeslets and derivatives between the foci $\xi =-c$ to $c$ given by
\begin{align}
Q_{ij}&=\int^{c}_{-c} d\xi G_{ij}(\bx-\xi \bp), \\
Q^D_{ij}&=\int^{c}_{-c} d\xi (c^2-\xi^2)G^D_{ij}(\bx-\xi \bp),\\
Q^R_{ijk}&=\int^{c}_{-c} d\xi (c^2-\xi^2)G^R_{ijk}(\bx-\xi \bp),\\
Q^{S}_{ijk}&=\int^{c}_{-c} d\xi (c^2-\xi^2)G^{S}_{ijk}(\bx-\xi \bp),\\
Q^Q_{ijk}&=\int^{c}_{-c} d\xi (c^2-\xi^2)^2G^Q_{ijk}(\bx-\xi \bp),
\end{align}
where $c^2=a^2(\lambda^2-1)/\lambda^2$.

Explicit expressions for spheroidal multipoles in terms of integrals of the stresslet, rotlet, and quadrupole are given in \citet{Einarsson2015b} and we add higher-order multipoles as needed here
\begin{align}
Q_{ij}&=\delta_{ij} I^0_1+x_ix_j  I^0_3 -(x_ip_j+x_jp_i) I^1_3+ p_ip_jI^2_3,\\
Q^D_{ij}&=2\delta_{ij}J^0_3+6 \Bigg[-x_ix_jJ^0_5+(x_ip_j+x_jp_i)J_5^1-p_ip_jJ^2_5\Bigg],\\
Q^R_{ijk}&=(\delta_{ik}x_j-\delta_{ij}x_k)J^0_3+(\delta_{ij}p_k-\delta_{ik}p_j)J^1_3,\\
Q^{S}_{ijk}&=\delta_{jk}x_iJ^0_3-\delta_{jk}p_iJ^1_3   \nonumber\\
                        &\quad+3 \Bigg[ -x_ix_jx_kJ^0_5+(x_ix_kp_j+x_jx_kp_i+x_ix_jp_k)J^1_5-(x_kp_ip_j+x_ip_jp_k+x_jp_ip_k)J^2_5+p_ip_jp_kJ^3_5\Bigg],\\
Q^Q_{ijk}&=6 \Bigg [-(\delta_{jk}x_i+\delta_{ik}x_j+\delta_{ij}x_k)K^0_5+(\delta_{jk}p_i+\delta_{ik}p_j+\delta_{ij}p_k)K^1_5 \Bigg]\nonumber\\
                       &\quad+30 \Bigg[ x_ix_jx_kK^0_7-(x_ix_kp_j+x_jx_kp_i+x_ix_jp_k)K^1_7 +(x_kp_ip_j+x_ip_jp_k+x_jp_ip_k)K^2_7-p_ip_jp_kK^3_7 \Bigg],\\                
Q^{R}_{ijk,m}&=(\delta_{ik}\delta_{jm}-\delta_{ij}\delta_{km})J^0_3 \nonumber\\
        & \quad+3(\delta_{ik}x_j-\delta_{ij}x_k)( p_m J^1_5-x_m J^0_5)\nonumber\\
          &  \quad+3(\delta_{ij}p_k-\delta_{ik}p_j)(p_m J^2_5-x_m J^1_5)\\
Q^{S}_{ijk,m}&=\delta_{jk}\delta_{im}J^0_3+3\delta_{jk}x_i (p_mJ^1_5-x_mJ^0_5) -3\delta_{jk}p_i(p_mJ^2_5-x_mJ^1_5) \nonumber\\
                        &\quad+3 \Bigg[ -(\delta_{im}x_jx_k+\delta_{jm}x_ix_k+\delta_{km}x_ix_j)J^0_5-5x_ix_jx_k(p_mJ^1_7-x_mJ^0_7) \nonumber\\
                        &\quad+(\delta_{im}x_kp_j+\delta_{km}x_ip_j  + \delta_{jm} x_kp_i+\delta_{km} x_jp_i    +  \delta_{im} x_jp_k+\delta_{jm}x_ip_k)J^1_5 \nonumber\\
                         &\quad+5(x_ix_kp_j+x_jx_kp_i+x_ix_jp_k)(p_mJ^2_7-x_mJ^1_7) \nonumber\\
                        &\quad-(\delta_{km}p_ip_j+\delta_{im}p_jp_k+\delta_{jm}p_ip_k)J^2_5  -5(x_kp_ip_j+x_ip_jp_k+x_jp_ip_k)(p_mJ^3_7-x_mJ^2_7)\nonumber\\
                       & \quad+5p_ip_jp_k(p_mJ^4_7-x_mJ^3_7)\Bigg],\\ 
Q^Q_{ijk,m}&=6 \Bigg [-(\delta_{jk}\delta_{im}+\delta_{ik}\delta_{jm}+\delta_{ij}\delta_{km})K^0_5-5(\delta_{jk}x_i+\delta_{ik}x_j+\delta_{ij}x_k)(p_mK^1_7-x_mK^0_7)\nonumber\\
                      &\quad+5(\delta_{jk}p_i+\delta_{ik}p_j+\delta_{ij}p_k)(p_mK^2_7-x_mK^1_7) \Bigg]\nonumber\\
                      &\quad+30 \Bigg[ (\delta_{im}x_jx_k+\delta_{jm}x_ix_k+\delta_{km}x_ix_j)K^0_7+7x_ix_jx_k(p_mK^1_9-x_mK^0_9)\nonumber\\
                     &\quad-(\delta_{im}x_kp_j+\delta_{km}x_ip_j+\delta_{jm}x_kp_i+\delta_{km}x_jp_i+\delta_{im}x_jp_k+\delta_{jm}x_ip_k)K^1_7\nonumber\\
                       &\quad-7(x_ix_kp_j+x_jx_kp_i+x_ix_jp_k)(p_mK^2_9-x_mK^1_9)\nonumber\\
                     &\quad+(\delta_{km}p_ip_j+\delta_{im}p_jp_k+\delta_{jm}p_ip_k)K^2_7 +7(x_kp_ip_j+x_ip_jp_k+x_jp_ip_k)(p_mK^3_9-x_mK^2_9)\nonumber\\
                     &\quad-p_ip_jp_kK^3_7 \Bigg],
 \end{align}
where
 \begin{align}\label{integI}
I^n_m &=\int^{c}_{-c} d\xi \frac{\xi^n}{|\bx-\xi \bp|^m},  \\   
J^n_m &=c^2 I^n_m -I^{n+2}_m, \label{integJ}\\ 
K^n_m&=c^2 J^n_m -J^{n+2}_m=c^4 I^n_m - 2 c^2I^{n+2}_m +I^{n+4}_m. \label{integK}
\end{align}

The integrals $I^n_{m}$ satisfy the relationship
 \begin{align}
&\frac{\partial}{\partial x_i} I^n_{m} = mp_i  I^{n+1}_{m+2} - mx_m I^{n}_{m+2}.
 \end{align}
To simplify integration one may employ an auxiliary coordinate system, $(x',y',z')$, with $\bp$ aligned with $x'$ such that
\begin{align}
& I^n_m=\int^c_{-c} d\xi \frac{\xi^n}{[(x'-\xi)^2+(y')^2+(z')^2]^{m/2}}=\int^c_{-c} d\xi \frac{\xi^n}{[(x'-\xi)^2+R^2]^{m/2}},
 \end{align}
where $R^2=(y')^2+(z')^2$. Integrating one obtains
 \begin{align}
&I^0_{-1}=\frac{1}{2} \left[R^2 \log \left(\frac{R_2-(x'-c)}{R_1-(x'+c)}\right)+c (R_1+R_2)-(R_2-R_1) x'\right],\\
&I^0_{1}=  \log \left(\frac{R_2-(x'-c)}{R_1-(x'+c)}\right) ,\\
&I^1_{1}= R_2-R_1+x' \log \left(\frac{R_2-(x'-c)}{R_1-(x'+c)}\right) ,\\
&I^2_{1}=  \frac{1}{2} \left[\left(2 x'^2-R^2\right) \log \left(\frac{R_2-(x'-c)}{R_1-(x'+c)}\right)+c (R_1+R_2)+3 (R_2-R_1) x'\right] ,\\
&I^0_{3}=  \frac{1}{R^2}\left[\frac{x'+c}{R_1}-\frac{x'-c}{R_2}\right],\\
&I^1_{3}=\frac{1}{R_1}-\frac{1}{R_2}+\frac{x'}{R^2} \left[ \frac{x'+c}{R_1}-\frac{x'-c}{R_2}\right],\\
&I^0_{5}=  \frac{1}{3 R^4}\left[\frac{\left(x'+c\right) \left(2 \left(x'+c\right)^2+3 R^2\right)}{R_1^3}-\frac{\left(x'-c\right) \left(2 \left(x'-c\right)^2+3 R^2\right)}{R_2^3}\right],\\
&I^1_{5}= \frac{1}{3 R^4} \Bigg[ \frac{\left(2 c^3+3 c R^2\right) x'+3 \left(2 c^2+R^2\right) x'^2+6 cx'^3+R^4+2 x'^4}{R_1^3} \nonumber\\
               &\hspace{1.5cm}  +\frac{\left(2 c^3+3 c R^2\right) x'-3 \left(2 c^2+R^2\right) x'^2+6 c x'^3-R^4-2 x'^4}{R_2^3}  \Bigg],\\
 &I^0_{7}=  \frac{1}{15 R^6} \Bigg[ \frac{\left(x'+c\right) \left[8 c^4+8 \left(4 c^3+5 c R^2\right) x'+4 \left(12 c^2+5 R^2\right) x'^2+20 c^2 R^2+32 cx'^3+15 R^4+8 x'^4\right]}{R_1^5} \nonumber\\
                  &\hspace{1.5cm} -\frac{\left(x'-c\right) \left[8 c^4-8 \left(4 c^3+5 c R^2\right) x'+4\left(12 c^2+5 R^2\right) x'^2+20 c^2 R^2-32 c x'^3+15 R^4+8 x'^4\right]}{R_2^5}\Bigg],\\
&I^1_{7}=\frac{1}{15 R^6}\Bigg[ \frac{1}{R_2^5} \Big[-3 R^6+\left(8 c^5+20 c^3 R^2+15 c R^4\right) x'-5 \left(8 c^4+12 c^2 R^2+3 R^4\right) x'^2\nonumber\\
                  &\hspace{2.5cm}+20 \left(4 c^3+3 c R^2\right) x'^3-20 \left(4 c^2+R^2\right) x'^4+40 c x'^5-8 x'^6\Big] \nonumber\\
                 &\hspace{1.5cm} -\frac{1}{R_1^5}\Big[-3 R^6-\left(8 c^5+20 c^3 R^2+15 c R^4\right) x'-5 \left(8 c^4+12 c^2 R^2+3 R^4\right) x'^2\nonumber\\
                 &\hspace{2.5cm}   -20 \left(4 c^3+3 c R^2\right) x'^3-20 \left(4 c^2+R^2\right)x'^4-40 cx'^5-8 x'^6\Big]\Bigg],
\end{align}
\begin{align}
&I^0_{9}= \frac{1}{35 R^8}\Bigg[\frac{1}{R_1^7}\Big[c \left(16 c^6+56 c^4R^2+70 c^2 R^4+35 R^6\right)+7 \left(16 c^6+40 c^4 R^2+30 c^2 R^4+5 R^6\right) x'\nonumber\\
                  &\hspace{2.5cm}+14 c \left(24 c^4+40 c^2 R^2+15 R^4\right) x'^2+70 \left(8 c^4+8 c^2 R^2+R^4\right) x'^3\nonumber\\
                   &\hspace{2.5cm}+280 c \left(2 c^2+R^2\right) x'^4+ 56\left(6 c^2+R^2\right) x'^5+112 c x'^6+16 x'^7\Big]\nonumber\\
               &\hspace{1.5cm}+\frac{1}{R_2^7}\Big[c \left(16 c^6+56 c^4 R^2+70 c^2 R^4+35 R^6\right)-7 \left(16 c^6+40 c^4 R^2+30 c^2 R^4+5 R^6\right) x'\nonumber\\
               &\hspace{2.5cm}+14 c \left(24 c^4+40 c^2 R^2+15 R^4\right) x'^2-70 \left(8 c^4+8 c^2 R^2+R^4\right) x'^3\nonumber\\
               &\hspace{2.5cm}+280 c \left(2 c^2+R^2\right) x'^4-56 \left(6 c^2+R^2\right) x'^5+112 c x'^6-16 x'^7 \Big]\Bigg],\\
 &I^1_{9}=\frac{1}{35 R^8}\Bigg[\frac{1}{R_1^7}\Big[5 R^8+\left(16 c^7+56 c^5 R^2+70 c^3 R^4+35 c R^6\right) x'+7 \left(16 c^6+40 c^4 R^2+30 c^2 R^4+5  R^6\right) x'^2 \nonumber\\
               &\hspace{2.5cm}+14 \left(24 c^5+40 c^3 R^2+15 c R^4\right) x'^3+70 \left(8 c^4+8 c^2 R^2+R^4\right) x'^4\nonumber\\
              &\hspace{2.5cm}+280 c \left(2 c^2+R^2\right) x'^5+56 \left(6 c^2+R^2\right) x'^6+112 c x'^7+16 x'^8 \Big]\nonumber\\
              &\hspace{1.5cm}+\frac{1}{R_2^7}\big[-5 R^8+\left(16 c^7+56 c^5 R^2+70 c^3 R^4+35 c R^6\right) x'-7 \left(16 c^6+40 c^4 R^2+30 c^2 R^4+5 R^6\right) x'^2\nonumber\\
              &\hspace{2.5cm} +14 \left(24 c^5+40 c^3 R^2+15 c R^4\right) x'^3-70 \left(8 c^4+8 c^2 R^2+R^4\right) x'^4\nonumber\\
              &\hspace{2.5cm}  +280 c \left(2 c^2+R^2\right) x'^5 -56 \left(6 c^2+R^2\right) x'^6+112 c x'^7-16 x'^8 \Big]\Bigg],
\end{align}
where on the surface of the particle we have
 \begin{align}
& R_1=\sqrt{\left(x'+c\right)^2+R^2},\nonumber\\
& R_2=\sqrt{\left(x'-c\right)^2+R^2},\nonumber\\
& R=\sqrt{\left(1-e^2\right) \left(a^2-x'^2\right)}.
 \end{align}

The integrals also satisfy the relationship
 \begin{align}
&I^n_{m} =  x' I^{n-1}_m+\frac{(n-1) I^{n-2}_{m-2}}{m-2}-\frac{c^{n-1}
   \left((-1)^{n} R_1^{2-m}+R_2^{2-m}\right)}{m-2}.
 \end{align}
Other integrals $J^n_m$, and $K^n_m$ can be calculated easily from equations \eqref{integJ} and \eqref{integK}. 

\section{A prolate spheroid in Stokes flow}
Following \citet{Einarsson2015b} we use the following ansatz for the disturbance flow field due to a prolate spheroid in a linear shear flow 
 \begin{align}
\label{spheroidastokes} 
u'_i &=Q^R_{ijk}\epsilon_{jkl} \Big[- \big\{A^R p_l p_m +B^R (\delta_{lm}-p_lp_m)\big\}\Omega'_m +C^R \epsilon_{lmn}p_mE^{\infty}_{no}p_o   \Big] \nonumber\\
                      &\quad +\big( Q^{S}_{ijk}+ \alpha Q^Q_{ijk} \big) \Big[ (A^S n^A_{jklm}+B^Sn^B_{jklm}+C^Sn^C_{jklm})E^{\infty}_{lm}+C^R (\epsilon_{jlm}p_kp_m+\epsilon_{klm}p_jp_m)\Omega'_l   \Big],
 \end{align}
where $A^R, B^R,C^R,A^S,B^S, C^S,$ and $\alpha$ are seven unknown scalar coefficients and are calculated by enforcing the no-slip boundary condition on the surface of the spheroid (see \citet{Einarsson2015b} for further details). Also  
 \begin{align}
& n^A_{jklm}=(p_jp_k-\frac{1}{3}\delta_{jk})(p_lp_m-\frac{1}{3}\delta_{lm}),\\
& n^B_{jklm}=p_jp_m\delta_{kl}+p_kp_m\delta_{jl}+p_jp_l\delta_{km}+p_kp_l\delta_{jm}-4p_jp_kp_lp_m,\\
& n^C_{jklm}=-\delta_{jk}\delta_{lm}+\delta_{jl}\delta_{km}+\delta_{kl}\delta_{jm} \nonumber\\
                        &\hspace{1.5cm}   +p_lp_m\delta_{jk}+p_jp_k\delta_{lm}-p_jp_m\delta_{kl}-p_kp_m\delta_{jl} -p_jp_l\delta_{km}-p_kp_l\delta_{jm}+p_jp_kp_lp_m.
 \end{align}

The constants for a prolate spheroid are
 \begin{align}
 &\alpha=\frac{1-e^2}{8e^2},\\
&A^R=\frac{e^2-1}{2 \mathcal{L} \left(e^2-1\right)+4 e},\\
&B^R=\frac{e^2-2}{2 \mathcal{L} \left(e^2+1\right)-4 e},\\
&C^R=-\frac{e^2}{2 \mathcal{L} \left(e^2+1\right)-4 e},\\
&A^S=-\frac{e^2}{2 \left(\mathcal{L} \left(e^2-3\right)+6 e\right)},\\
&B^S=\frac{e^2 \left(\mathcal{L} \left(e^2-1\right)-4 e^3+2 e\right)}{4 \left(-3 \mathcal{L} \left(e^2-1\right)+4 e^3-6 e\right) \left(\mathcal{L} \left(e^2+1\right)-2e\right)},\\ 
& C^S=\frac{e^2-e^4}{3 \mathcal{L} \left(e^2-1\right)^2+2 e \left(5 e^2-3\right)},\\ 
&\mathcal{L}=\log \left(-\frac{e+1}{e-1}\right).
 \end{align}
 
 \subsection{Jeffery Orbits} 
The torque on a spheroid can be calculated by linearly superposing the contributions from all the contained rotlets, $u_i=Q^R_{ijk}\epsilon_{jkl} B_l$, as
  \begin{align}
 &\label{torqueprolate} \bL_0=-16\pi \int^{c}_{-c} (c^2-\xi^2) d \xi \bB = -\frac{64 \pi c^3}{3} \bB,
 \end{align}
 where from \eqref{spheroidastokes} the rotlet strength is
  \begin{align}
 &\bB= -\big\{A^R \bp\bp +B^R (\bI-\bp\bp)\big\} \cdot \bOmega'_0 +C^R \bp \times (\bE^{\infty}\cdot \bp).
 \end{align}
 Ultimately, because the torque on the body is zero then the rotlet strength must be zero and hence $\bB=\bzero$. The Jeffery orbit solution immediately follows as this requires that
\begin{align*}
\bOmega_0 =\bOmega^{\infty}+\Lambda\, \bp \times \bE^{\infty} \cdot \bp,
\end{align*}
where $\Lambda = C^R/B^R$.

The disturbance flow field may now be written more simply as
 \begin{align}
u_{0i}=A^{\infty}_{ij}x_j +\big( Q^{S}_{ijk}+ \alpha Q^Q_{ijk} \big) D_{jklm}E^{\infty}_{lm},
 \end{align}
 where
  \begin{align}
  D_{jklm}= A^S n^A_{jklm}+B^Sn^B_{jklm}+C^Sn^C_{jklm}+ C^R\Lambda (\epsilon_{jqt}p_kp_t+\epsilon_{kqt}p_jp_t) \epsilon_{pql}p_pp_m.
 \end{align}
 
From this solution of the Newtonian disturbance flow, the strain-rate tensor maybe be calculated
 \begin{align}\label{strainraten1}
 \dot{\bgamma}_{0} &= \bM : \bE^{\infty},
 \end{align}
 where
 \begin{align}\label{strainraten2}
 M_{imsl} &=  2 \delta_{il} \delta_{ms} +\Big( Q^{ST}_{ijkm}+ \alpha Q^{QT}_{ijkm} \Big) D_{jkls},
 \end{align}
and we've defined $Q^{ST}_{ijkm}=Q^{S}_{ijk,m}+Q^{S}_{mjk,i}$, and $Q^{QT}_{ijkm}=Q^{Q}_{ijk,m}+Q^{Q}_{mjk,i}$.

\section{Shear-thinning correction}\label{A:E:NN}
In order to calculate the correction in the orientational dynamics of the prolate spheroid due to  shear thinning  given by \eqref{omega1},
\begin{align*}
\bOmega_1 &=\frac{1}{2}(1-\beta)(1-n)\bRh_{\bL\bOmega}^{-1}\cdot\int_{\fV}|\dot{\gamma}_0|^2 \dot{\bgamma}_0:\bEh_{\bOmega}\d V,
\end{align*}
we need both the strain-rate tensor of the Newtonian solution,$\dot{\bgamma}_0$, given by \eqref{strainraten1} and the operators $\bRh_{\bL\bOmega}^{-1}$ and $\bEh_{\bOmega}$ from the rigid-body motion problem that we now show.
 \subsection*{Rigid-body motion}
The flow field due to a prolate spheroid rotating with $\hat{\bOmega}$ in a quiescent flow, obtained from \eqref{spheroidastokes}, is 
  \begin{align}
&\hat{u}_{i}=\left[-\epsilon_{jkl} Q^R_{ijk} \lb\{A^R p_l p_s +B^R (\delta_{ls}-p_lp_s)\rb\} +\big( Q^{S}_{ijk}+ \alpha Q^Q_{ijk} \big)  C^R (\epsilon_{jsm}p_kp_m+\epsilon_{ksm}p_jp_m) \right] \hat{\Omega}_{s}.
 \end{align}

The strain-rate tensor is then
   \begin{align}
\hat{\dot{\bgamma}} &= 2 \hat{\bE}_{\bOmega} \cdot \hat{\bOmega},
\end{align}
where
\begin{align} 
\hat{E}_{\Omega_{ims}} &=\frac{1}{2} \Big[ -\epsilon_{jkl} Q^{RT}_{ijkm} \lb\{A^R p_l p_s +B^R (\delta_{ls}-p_lp_s)\rb\} +\big( Q^{ST}_{ijkm}+ \alpha Q^{QT}_{ijkm} \big) C^R (\epsilon_{jsm}p_kp_m+\epsilon_{ksm}p_jp_m)\Big],
 \end{align}
and $Q^{RT}_{ijkm}=Q^{R}_{ijk,m}+Q^{R}_{mjk,i}$. 
 
 The torque exerted on the particle can be found by integrating the rotlet density such that
   \begin{align}\label{torqueprolate:auxiliary} 
   \hat{\bL}&=16\pi \int^{c}_{-c} d \xi(c^2-\xi^2)     \lb(A^R \bp\bp +B^R (\bI-\bp\bp)\rb) \cdot \hat{\bOmega},\nonumber\\
 & = \frac{64 \pi c^3}{3}  \lb(A^R \bp\bp +B^R (\bI-\bp\bp)\rb) \cdot \hat{\bOmega}.
 \end{align}
Using the definition of the resistance, $\bLh=-\hat{\bR}_{\bL \bOmega} \cdot \hat{\bOmega}$, we obtain
  \begin{align}
\hat{\bR}_{\bL \bOmega}^{-1}=-\frac{3}{64 \pi c^3}\lb[\frac{1}{A^R}\bp\bp +\frac{1}{B^R}(\bI-\bp\bp)\rb].
\end{align}

%
%
%
%
%
%
%
%
%
%
%
%
%
%
%
%
%
%
%
%
%
%
%
%
%
%
%
%
%
%
%
%
%


\end{document}

%% file: definitions.tex
\DeclareMathAlphabet\mathcalbf{OMS}{cmsy}{b}{n}

\def \t{\tensorsym}
\def \lb{\left}
\def \rb{\right}

\def \d{\,\text{d}}

\def \bgamma{\boldsymbol{\gamma}}

\def \bgammad{\dot{\boldsymbol{\gamma}}}
\def \bgammadh{\hat{{\dot{\bgamma}}}}
\def \gammad{{\dot{\gamma}}}

\def \bnabla{\boldsymbol{\nabla}}

\def \bOmega{\mathbf{\Omega}}

\def \phid{\dot{\phi}}

\def \bsigmah{\hat{\boldsymbol{\sigma}}}

\def \btau{\boldsymbol{\tau}}

\def \bzero{\mathbf{0}}

\def \bA{\mathbf{A}}

\def \bB{\mathbf{B}}

\def \fB{\mathcal{B}}

\def \bE{\mathbf{E}}
\def \bEh{\hat{\mathbf{E}}}

\def \tEh{\mathsf{\t{\hat{E}}}}

\def \tEh{\mathsf{\t{\hat{E}}}}

\def \bF{\mathbf{F}}

\def \tF{\mathsf{\t F}}

\def \bG{\mathbf{G}}

\def \bI{\mathbf{I}}

\def \bL{\mathbf{L}}
\def \bLh{\hat{\mathbf{L}}}

\def \bM{\mathbf{M}}

\def \fO{\mathcal{O}}

\def \bR{\mathbf{R}}
\def \bRh{\hat{\mathbf{R}}}

\def \tRh{\mathsf{\t{\hat{R}}}}

\def \bT{\mathbf{T}}
\def \bTh{\hat{\mathbf{T}}}

\def \tTh{\mathsf{\t{\hat{T}}}}

\def \bU{\mathbf{U}}

\def \tU{\mathsf{\t U}}
\def \tUh{\mathsf{\t{\hat{U}}}}

\def \fV{\mathcal{V}}

\def \be{\mathbf{e}}

\def \bn{\mathbf{n}}

\def \bp{\mathbf{p}}
\def \bpd{\dot{\bp}}

\def \bu{\mathbf{u}}

\def \bx{\mathbf{x}}